\documentclass[aps,prx,reprint,groupedaddress,amsmath,amssymb,superscriptaddress]{revtex4-2}
\usepackage{graphicx}
\usepackage{float}
\usepackage{multirow}
\usepackage{natbib,twoopt}
\usepackage[breaklinks=true]{hyperref}
\usepackage{xcolor}
\usepackage{amssymb}
\usepackage{stackengine,graphicx}
\usepackage{amsmath}
\usepackage{hyperref}
\usepackage{color}
\usepackage{stackengine}
\usepackage{subfigure}
\usepackage{verbatim}
\usepackage{bbold}
\usepackage{stmaryrd} 
\usepackage{overpic}

\newcommand{\df}[2]{\frac{\partial #1}{\partial #2}}
\newcommand{\ds}[2]{\frac{\partial^2 #1}{\partial #2^2}}
\newcommand{\eps}{\varepsilon}

\newcommand{\bas}{{\bf e}_z}
\renewcommand{\o}{\omega}


\begin{document}

\title{Probing ultrafast magnetization dynamics via synthetic axion fields}

\author{Leon Shaposhnikov}
\email{These authors have equally contributed to this work}
\affiliation{School of Physics and Engineering, ITMO University, Saint  Petersburg 197101, Russia}

\author{Eduardo Barredo-Alamilla}
\email{These authors have equally contributed to this work}
\affiliation{School of Physics and Engineering, ITMO University, Saint  Petersburg 197101, Russia}

\author{Frank Wilczek}
\affiliation{Department of Physics, Stockholm University, Stockholm 10691, Sweden}
\affiliation{Center for Theoretical Physics, Massachusetts Institute of Technology, Cambridge, Massachusetts 02139, USA}
\affiliation{Department of Physics, Arizona State University, Tempe, Arizona 25287, USA}
\affiliation{Wilczek Quantum Center, Department of Physics and Astronomy, Shanghai Jiao Tong University, Shanghai 200240, China}

\author{Maxim A. Gorlach}
\email{m.gorlach@metalab.ifmo.ru}
\affiliation{School of Physics and Engineering, ITMO University, Saint  Petersburg 197101, Russia}

\begin{abstract}
Spatial structuring of materials at subwavelength scales underlies the concept of metamaterials possessing exotic properties beyond those of the constituent media. Temporal modulation of material parameters enables further functionalities. Here, we show that high-frequency oscillations of spatially uniform magnetization generate an effective dynamic axion field embedding the amplitude and phase of magnetization oscillations. This allows one to map ultrafast magnetization dynamics using a probe signal with much lower frequency. 
\end{abstract}

\maketitle


\section{Introduction}

The optics of time-varying media~\cite{Caloz2020,Caloz2020a,Galiffi2022} has a long history with pioneering studies dating back to 1950s-1970s~\cite{Morgenthaler1958,Felsen1970}. Recent  advances in materials engineering and nanofabrication have revived interest in this field, bringing the realization of time-modulated photonic structures within experimentalists' reach~\cite{Saha2023,Moussa2023}. 

Modulating the material parameters in time unlocks a set of interesting functionalities~\cite{Engheta2023}. Since the modulation breaks time translation symmetry, energy is generally not conserved~\cite{Morgenthaler1958}.  It enables strong and selective amplification of radiation~\cite{Lyubarov2022,Wang2023}, frequency conversion and even radiation from stationary charges~\cite{Shapiro2023}. Furthermore, temporal modulation of the medium can break time-reversal symmetry $\mathcal{T}$ at optical frequencies even in the absence of static magnetic fields  paving a way towards strongly non-reciprocal optical structures~\cite{Wang2020,Cardin2020}. These possibilities have stimulated much work, as reflected in recent reviews~\cite{Caloz2020,Caloz2020a,Galiffi2022}.

\begin{figure}[b]
	\centering
	\includegraphics[width=0.45\textwidth]{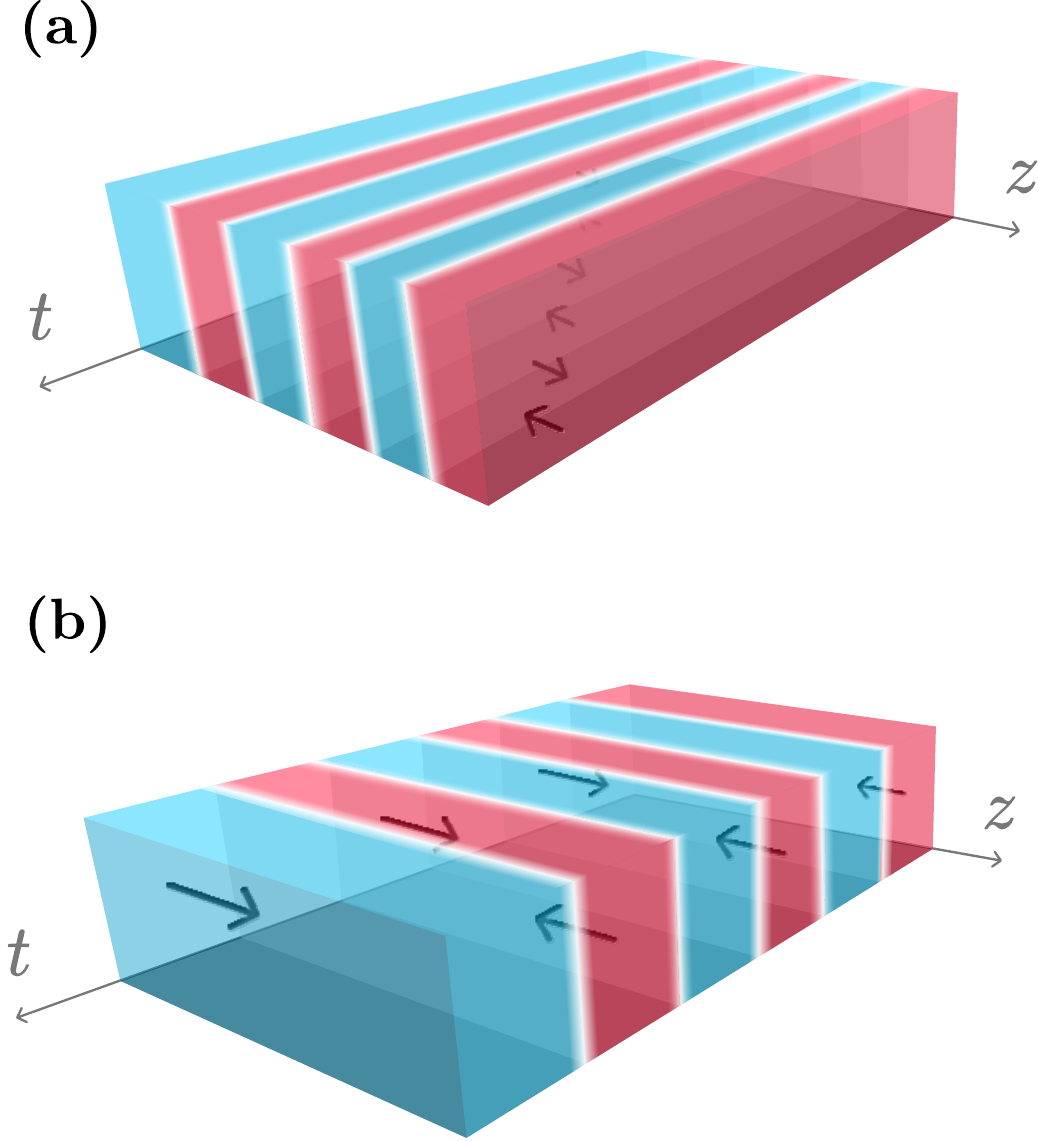}
	\caption{Illustration of a parallel between photonic crystals structured in space (a) and photonic time crystals modulated in time (b).
 }
	\label{fig:structure}
\end{figure}

The physics of time-varying media exhibits parallels to the related area of photonic crystals~--- artificial structures with engineered spatial periodicity. Similar to how emergent properties of photonic crystals originate from their spatial structuring, the physics of time-modulated media is rooted in the specific form of material parameters temporal modulation (Fig.~\ref{fig:structure}). For that reason, materials periodically modulated in time are often called {\it photonic time-crystals (PTCs)}. Note that these structures break translation symmetry in time due to the external stimulation, which distinguishes them from time crystals proper ~\cite{Shapere2012,Wilczek2012}, where the $\mathcal{T}$-breaking is spontaneous.

While PTCs often break $\mathcal{T}$ symmetry and reciprocity, the variety of available nonreciprocal responses remains largely unexplored.  Axion electrodynamics~\cite{Wilczek1987}, which has been the focus of much interest in fundamental physics~\cite{Millar2023}, condensed matter~\cite{Wu2016,Nenno2020,Sekine2021,Xu2023} and photonics~\cite{Shaposhnikov2023,Asadchy2024,Silveirinha2023}, is an outstanding target in this area.

Emergent axion fields demonstrated previously in condensed matter and photonic systems are largely static. However, the excitation of spin waves (magnons) with a suitable symmetry renders effective axion field time-dependent~\cite{Li2010,Private}. Coupling between magnons and electromagnetic field results in axion-polaritons~\cite{Li2010}, and up to now the experimental effort was to probe the dispersion of those hybrid light-matter particles~\cite{Li2010}.

In this Article, we approach the problem of dynamic axion fields from the different perspective assuming that the oscillations of magnetization are not monochromatic, but start at a well-defined moment of time. We demonstrate that these ultrafast oscillations enable an effective axion field. Surprisingly, its magnitude depends not only on the amplitude of oscillations, but also on their initial phase. In turn, the effective axion field affects the propagation of light, and ultrafast magnetization dynamics is imprinted in the cross-polarized time-reflection and time-transmission coefficients of probe waves with much lower frequency.



Here we will give qualitative arguments supporting this picture, develop an approximate analytical description, and finally validate it through comparison with rigorous numerical simulations.

\section{Time boundaries and synthetic axion field}

The simplest case of temporal variation is an abrupt change of the medium material parameters [Fig.~\ref{fig:structure}]. We call this situation {\it time boundary} in analogy to the spatial boundaries separating the domains with distinct properties. The change of the material parameters at time boundary generates a singularity in time derivatives of the fields in Maxwell's equations
\begin{equation}\label{eq:Maxwell1}
\nabla\times {\bf E}=-\frac{1}{c}\,\df{{\bf B}}{t}\:,\mspace{10mu} \nabla\times {\bf H}=\frac{1}{c}\,\df{{\bf D}}{t}\:,
\end{equation}
where, typically, ${\bf D}=\hat{\eps}\,{\bf E}$ and ${\bf H}=\hat{\mu}^{-1}\,{\bf B}$. Integrating Eqs.~\eqref{eq:Maxwell1} across a small interval of time including the time boundary $t_0$, we recover the continuity of the fields ${\bf B}$ and ${\bf D}$ at a given moment of time in all points of space:
\begin{equation}\label{eq:BC1}
    {\bf B}_2({\bf r},t_0)={\bf B}_1({\bf r},t_0)\:,\mspace{10mu} {\bf D}_2({\bf r},t_0)={\bf D}_1({\bf r},t_0)\:,
\end{equation}
where subscripts $1$ and $2$ denote the fields before and after the time boundary. On the other hand, the presence of a spatially homogeneous time-varying axion field $\chi$ modifies the equations of electromagnetism yielding
\begin{equation}
\nabla\times{\bf E}=-\frac{1}{c}\,\df{{\bf B}}{t}\:,\mspace{10mu} \nabla\times{\bf H}=\frac{1}{c}\,\df{{\bf D}}{t}+\frac{1}{c}\,\df{\chi}{t}\,{\bf B}\:.    
\end{equation}
Typically, $\chi$ is of the order of $10^{-3}$ for the existing condensed matter systems~\cite{Nenno2020} increasing up to $10^{-2}-10^{-1}$ for specially designed metamaterials~\cite{Shaposhnikov2023,Asadchy2024}. 

A spatially and temporally homogeous axion field has no effect on electrodynamics. However, an abrupt change of the material parameters can modify $\chi$ and render the derivative $\partial\chi/\partial t$ singular. As a result, the conditions at a time boundary in axion electrodynamics  read:
\begin{gather}
{\bf B}_2={\bf B}_1\:,\label{eq:BC2a}\\
{\bf D}_2+\chi_2\,{\bf B}_2={\bf D}_1+\chi_1\,{\bf B}_1\:,\label{eq:BC2b}
\end{gather}
where $({\bf r},t_0)$ arguments are omitted for brevity. The change in the boundary condition Eq.~\eqref{eq:BC2b} provides a signature for the effective axion field. 

As recently suggested, the metamaterial composed of layers with alternating out-of-plane magnetization exhibits an isotropic axion response~\cite{Shaposhnikov2023}. We therefore can anticipate that temporal modulation of spatially uniform magnetization will enable effective axion field as well. Moreover, if the modulation is not periodic, the resulting axion field can be time-dependent.

\section{Derivation of the effective axion field}


Let us support that intuition analytically.
To reveal the origins of the effective axion response, we consider an unbounded gyrotropic medium with the scalar permittivity $\eps$ and inverse permeability tensor of the form
\begin{equation}\label{eq:InvMu}
    \hat{\mu}^{-1}=
    \begin{pmatrix}
     \mu^{-1} & i\,g(t) & 0\\
     -i\,g(t) & \mu^{-1} & 0\\
     0 & 0 & \mu^{-1}
    \end{pmatrix}\:,
\end{equation}
where off-diagonal component $g(t)$ is responsible for $\mathcal{T}$-breaking and arises due to the medium magnetization.

We assume that the temporal modulation switches on at the moment of time $t=0$ and since then the gyrotropy $g(t)$ experiences a periodic modulation:
\begin{equation}
    g(t)=\sum\limits_{n\not=0}\,g_n\,e^{-in\Omega\,t}\:,
\end{equation}
while the permittivity $\eps$ and permeability $\mu$ of the medium stay constant. To isolate the effects of effective axion field from the conventional gyrotropy, we set the average magnetization to zero: $g_0=0$.

In this analysis, the modulation frequency $\Omega$ is much larger than the spatial frequency $c\,k$ of the wave propagating in the medium. In the other words, the ratio $\xi=c\,k/\Omega$ plays the role of a small parameter. Such rapid modulation limit of photonic time crystal is analogous to the metamaterial regime [Fig.~\ref{fig:structure}] when the spatial period of structuring is much smaller than the wavelength.

Since the medium is spatially homogeneous, the wave vector ${\bf k}$ of the plane wave is conserved, and hence the magnetic field can be expanded in Floquet series as
\begin{equation}\label{eq:Floquet}
    {\bf B}({\bf r},t)=e^{i{\bf k}\cdot{\bf r}-i\omega\,t}\,\sum\limits_{n}\,{\bf B}_n\,e^{-in\Omega\,t}\:,
\end{equation}
the similar expansions are valid for the fields ${\bf E}$, ${\bf D}$ and ${\bf H}$. Equations~\eqref{eq:Maxwell1} yield the wave equation for the magnetic field:
\begin{equation}\label{eq:WaveEq}
    \frac{\eps}{c^2}\,\ds{{\bf B}}{t}+\nabla\times\nabla\times\left(\hat{\mu}^{-1}\,{\bf B}\right)=0\:.
\end{equation}
For clarity, we first analyze the scenario when the wave vector ${\bf k}$ is aligned with the direction of magnetization, $Oz$. Combining Eqs.~\eqref{eq:Floquet},\eqref{eq:WaveEq}, we derive Floquet harmonics of magnetic field for $n\not=0$:
\begin{equation}\label{eq:Bn}
    {\bf B}_n=-\frac{i\xi^2}{\eps\,n^2}\,g_n\,\left[\bas\times{\bf B}_0\right]+O(\xi^4)\:,
\end{equation}
Next, using Eq.~\eqref{eq:Maxwell1}, we derive the Floquet harmonics of electric displacement, where the leading-order term reads:
\begin{equation}\label{eq:Dn}
    {\bf D}_n=-\frac{i\xi}{n}\,g_n\,{\bf B}_0+O(\xi^3)\:.
\end{equation}
The microscopic fields ${\bf B}$ and ${\bf D}$ satisfy the conventional boundary conditions Eq.~\eqref{eq:BC1} at the temporal boundary $t=0$:
\begin{gather*}
    \left.{\bf B}\right|_{t=-0}=\sum\limits_{n}\,{\bf B}_n\:,\mspace{10mu}
    \left.{\bf D}\right|_{t=-0}=\sum\limits_{n}\,{\bf D}_n\:.
\end{gather*}

From the practical perspective, it is convenient to describe time-modulated medium in terms of the averaged fields ${\bf B}_0$ and ${\bf D}_0$ excluding rapidly oscillating higher-order harmonics. Using Eqs.~\eqref{eq:Bn},\eqref{eq:Dn} and keeping the terms up to the first power in $\xi$, we recover the following conditions at the time interface:
\begin{equation}\label{eq:BC3}
  \left.{\bf B}\right|_{t=-0}={\bf B}_0\:,\mspace{10mu} 
  \left.{\bf D}\right|_{t=-0}={\bf D}_0+\chi\,{\bf B}_0\:,
\end{equation}
where
\begin{equation}\label{eq:EffAxion}
    \chi=-i\xi\,\sum\limits_{n\not=0}\,\frac{g_n}{n}\:.
\end{equation}
The obtained boundary conditions coincide with those in the axion electrodynamics [Eqs.~\eqref{eq:BC2a},\eqref{eq:BC2b}] and therefore the coefficient $\chi$ above is the effective axion field generated by the rapid modulation of the medium. Equivalently, this result can be recast in the form
\begin{equation}\label{eq:EffAxion2}
    \chi=-\frac{\xi\,\Omega}{T}\,\int\limits_{t_0-T/2}^{t_0+T/2}\,g(t)\,\left(t-t_0\right)\,dt\:,
\end{equation}
where $T=2\pi/\Omega$ is the modulation period. 

The effective axion field appears to be sensitive to the phase of the temporal modulation. For instance, if gyrotropy is modulated in time as $g(t)=g_0\,\sin\left(\Omega t+\varphi\right)$, the phase $\varphi$ will be imprinted in the effective axion field:
\begin{equation}\label{eq:PhaseDependence}
    \chi=\xi\,g_0\,\cos\varphi\:.
\end{equation}
%
Therefore, the phase of ultrafast magnetization oscillations determines the effective axion field, while the latter governs the propagation of electromagnetic waves with much lower frequency. This provides a useful probe of ultrafast magnetization dynamics.

Above, we analyzed the situation when $g(t)$ is a periodic function of time which yields time-independent axion field. However, our analysis can be readily generalized towards non-periodic functions $g(t)$ which rapidly oscillate in time gradually changing the oscillation amplitude. In such case, $\chi$ is promoted to the dynamic field proportional to the envelope function of rapid magnetization oscillations, Eq.~\eqref{eq:EffAxion2}.

The amplitude of the effective axion field depends on the angle $\theta$ between the wave vector ${\bf k}$ and the magnetization axis $Oz$ scaling as $\cos\theta$ (see Supplementary Materials~\cite{Supplement}). This is the manifestation of spatial dispersion effects ubiquitous in photonic time crystals~\cite{Rizza2022}. To maximize the effect,  we analyze the situation when the wave vector of the probe wave is parallel to the direction of magnetization.

\section{Numerical validation}

\begin{figure}[t]
  \centering
  \includegraphics[width=0.47\textwidth]{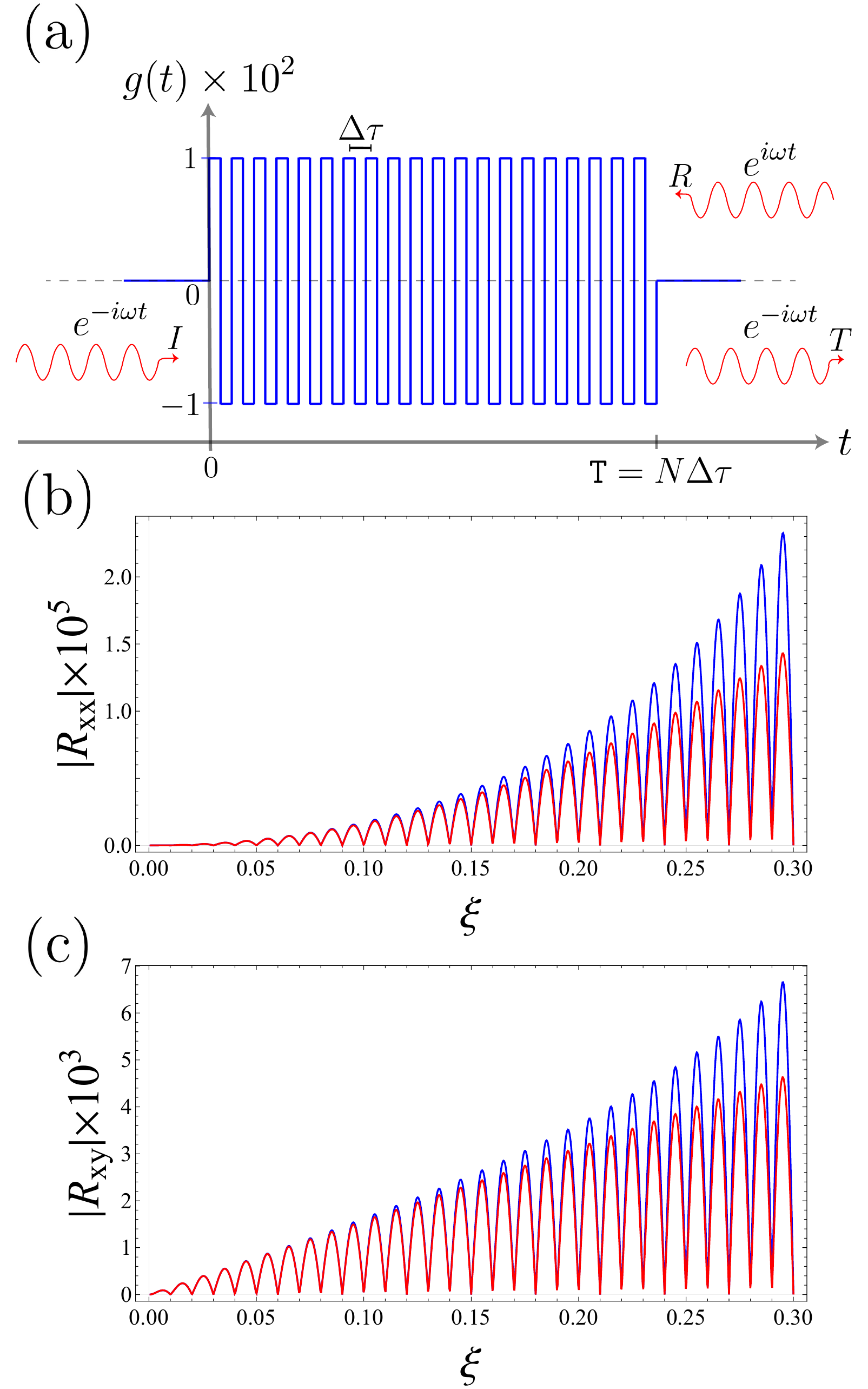}
  \caption{(a) Modulation of the magnetization $g(t)$ with the period $\Delta\tau$ for the PTC composed of $N$ modulation periods. The average magnetization vanishes. (b,c) Scattering characteristics of the time-modulated medium. Red and blue lines depict effective medium approximation and transfer matrix method, respectively. (b) and (c) represent co- and cross-polarized reflection. For the red lines $g=0.01$, $\varepsilon_0 = \mu_0 = \mu =\varepsilon= \varepsilon_{\text{eff}} = 1$,  $\mu_\text{eff} =  \mu\left( 1-\frac{\pi^2}{12}\frac{|g_0|^2}{\eps\mu} \xi^2\right)$, and $\chi_{\text{eff}}$ is given by Eq. (\ref{effectiveaxionfieldg0}), $N=50$.}
\label{Fig:validationeffectiveaxionslab}
\end{figure}

Next we examine the impact of the effective axion field on the propagation of a probe wave. For clarity, we consider first an artificial situation when the modulation of the medium magnetization $g(t)$ is realized in a stepwise manner with a time interval $\Delta\tau/2$ between the consecutive magnetization flips, see Fig.~\ref{Fig:validationeffectiveaxionslab}(a). 

Such modulation creates a set of temporal boundaries, each of them breaks the invariance of the system in time keeping the translational invariance intact. As a result, the wave vector ${\bf k}$ of impinging wave is conserved, while the frequency $\o$ is altered to satisfy the dispersion relation.

We assume that after $N$ modulation periods the parameters of the medium are returned to their original values [Fig.~\ref{Fig:validationeffectiveaxionslab}(a)]. As a result, the dispersion relation guarantees that the frequency of the wave is either $-\o$ or $\o$, which is called time-transmitted and time-reflected waves, respectively. Physically, time-reflected wave corresponds to the field configuration propagating in $-{\bf k}$ direction with a conjugated amplitude.

To find the amplitudes of reflected and transmitted waves after multiple modulation periods, we employ the transfer matrix method in time domain. Building on the continuity of the fields ${\bf B}$ and ${\bf D}$ at a time boundary [Eq.~\eqref{eq:BC1}], this method introduces a transfer matrix that relates the amplitudes of the fields at different times, before $(i)$ and after $(j)$ the magnetization flip~\cite{Supplement}:

\begin{align}
    \begin{pmatrix}
        \vphantom{\dfrac{Y_{i,\eta}}{Y_{j,\eta}}} B_{j,\eta} \\
        \vphantom{\dfrac{Y_{i,\eta}}{Y_{j,\eta}}} B'_{j,\eta}
    \end{pmatrix} & =  \frac{1}{2}
    \begin{pmatrix}
        1 + \dfrac{Y_{i,\eta}}{Y_{j,\eta}} &  1 - \dfrac{Y_{i,\eta}}{Y_{j,\eta}} \\
        1 - \dfrac{Y_{i,\eta}}{Y_{j,\eta}} &  1 + \dfrac{Y_{i,\eta}}{Y_{j,\eta}}
    \end{pmatrix}
    \begin{pmatrix}
       \vphantom{\dfrac{Y_{i,\eta}}{Y_{j,\eta}}} B_{i,\eta} \\
       \vphantom{\dfrac{Y_{i,\eta}}{Y_{j,\eta}}} B'_{i,\eta}
    \end{pmatrix},
\end{align}
where $B_{\alpha,\eta}$ and $B'_{\alpha,\eta}$ are the amplitudes of the forward and backward waves respectively. The indices $\alpha = i,j= \rightarrow,\leftarrow$ indicate the time-layer with positive or negative gyrotropy, $\eta = \pm 1$ corresponds to the right- and left-handed modes, $Y_{\alpha, \eta} = \sqrt{\varepsilon_\alpha/ \mu_{\alpha,\eta}}$ is the admittance within the time-layer with permeability and permittivity  $\mu_{\alpha,\eta} = \mu^2 /(\mu -\eta g_\alpha)$ and $\varepsilon_\alpha=\varepsilon$. Multiple time boundaries result in the multiplication of those matrices. 

Applying this technique, we calculate the amplitudes of time-transmitted and time-reflected waves for the chosen stepwise magnetization modulation and compare the results with the effective medium prediction. The latter assumes no gyrotropy, as it is averaged down to zero, but introduces an effective axion field
\begin{align}
    \chi_\text{eff} & = \frac{\pi}{2} \, \frac{g_0}{ \mu_\text{eff}^2} \,\xi,
\label{effectiveaxionfieldg0}\end{align}
where  $g_0$ is the gyrotropy of the initial layer of the PTC's unit-cell, $\xi =  c\,\Delta\tau /\lambda$ is the period-to-wavelength ratio, and $\lambda$ is the vacuum wavelength. The effective permittivity corresponds to that of the constituent layers: $\varepsilon_{\text{eff}}=\varepsilon $, while the permeability is renormalized as $\mu_\text{eff} =  \mu\left( 1-\frac{\pi^2}{12}\frac{|g_0|^2}{\eps\mu} \xi^2\right)$~\cite{Supplement}.

Time-transmitted and time-reflected fields are related to the incident ones via $\mathbf{B}^t = \hat{T}\, \mathbf{B}^i$ and $\mathbf{B}^r = \hat{R} \, \mathbf{B}^i$. In the effective medium approach, the reflection and transmission matrices $\hat{R}$ and $\hat{T}$ are found analytically as~\cite{Supplement}
\begin{align}
    T_{xx} = & \hphantom{-}\; T_{yy}  = \cos\psi - \frac{i}{2} A_+(\chi_{\text{eff}}) \sin\psi, \label{TxxAxion} \\
    T_{xy}  = & -T_{yx}  = 0, \label{TxyAxion}\\
    R_{xx}  = & \hphantom{-} \; R_{yy}  =  \frac{i}{2} A_-(\chi_{\text{eff}}) \sin\psi, \label{RxxAxion}\\
    R_{xy}  = & - R_{yx}  = -i \sqrt{\dfrac{\mu_{\text{eff}} }{\varepsilon_{\text{eff}}}} \chi_{\text{eff}} \sin\psi\:, \label{RxyAxion}
\end{align}
where $A_{\pm}(\chi)=\left(\sqrt{\frac{\mu_0 \varepsilon_{\text{eff}}}{\varepsilon_0 \mu_{\text{eff}}}} \pm \sqrt{\frac{\varepsilon_0 \mu_{\text{eff}}}{\mu_0\varepsilon_{\text{eff}}}} + \sqrt{\frac{\mu_0 \mu_{\text{eff}}}{\varepsilon_0\varepsilon_{\text{eff}}}}\chi^2 \right)$, $\mu_0$ $ (\varepsilon_0)$ and  $\mu_{\text{eff}} $ $(\varepsilon_{\text{eff}})$ are the permeability (permittivity) of unmodulated and modulated medium, respectively. $\psi = 2\pi N \xi /\sqrt{\mu_{\text{eff}} \varepsilon_{\text{eff}}} $ is the temporal path inside the axion slab and $N$ represents the number of modulation periods.

Equations \eqref{TxxAxion}-\eqref{RxyAxion} suggest that the cross-polarized time-transmission vanishes, while cross-polarized time-reflection provides a smoking gun of the effective axion field.

Furthermore, denoting $|T|^2=|T_{xx}|^2+|T_{yx}|^2$ and $|R|^2=|R_{xx}|^2+|R_{yx}|^2$, we observe that $|T|^2-|R|^2=1$, while $|T|^2+|R|^2 \neq 1$. This can be interpreted as conservation of momentum of the incident wave, but non-conservation of its energy, as further discussed in~\cite{Supplement}.

To examine the limits of the effective description, we fix the number of the modulation periods $N=50$  varying the ratio between the modulation period and the period of the wave, $\xi$. In the simulations presented in Fig.~\ref{Fig:validationeffectiveaxionslab}(b-c) this ratio $\xi$ varies in the range from almost 0 to 0.3.

We observe that for rapid modulation with $\xi \ll 1$, the rigorous results for the time-reflected signal (blue lines) closely match the prediction of the effective medium approach (red lines) thus justifying the terminology of effective axion field. As $\xi$ increases up to $0.2$, the two approaches still agree qualitatively, but quantitative differences become noticeable.



The above situation can be readily generalized towards non-periodic magnetization modulation. As an illustrative example, we analyze the linear increase of the modulation amplitude resulting in a linearly growing axion field $\chi = b_0 t$, as in Weyl semimetals with non-zero chiral chemical potential~\cite{Zyuzin2012Apr,Zyuzin2012Sep,Guo2023,Gomez2024Mar} where the chiral magnetic effect can be observed~\cite{Fukushima2008Oct,Li2016Jun}. As we demonstrate~\cite{Supplement}, this enables optical phenomena characteristic to  Carroll-Field-Jackiw electrodynamics~\cite{Carroll1990Feb} which was a subject of recent investigations~\cite{Casana2008Jul,Qiu2017Feb,Chen2019Feb,Silva2020Oct}.



\section{Mapping magnetization dynamics}

\begin{figure*}
  \centering
	\includegraphics[width=0.9\textwidth]{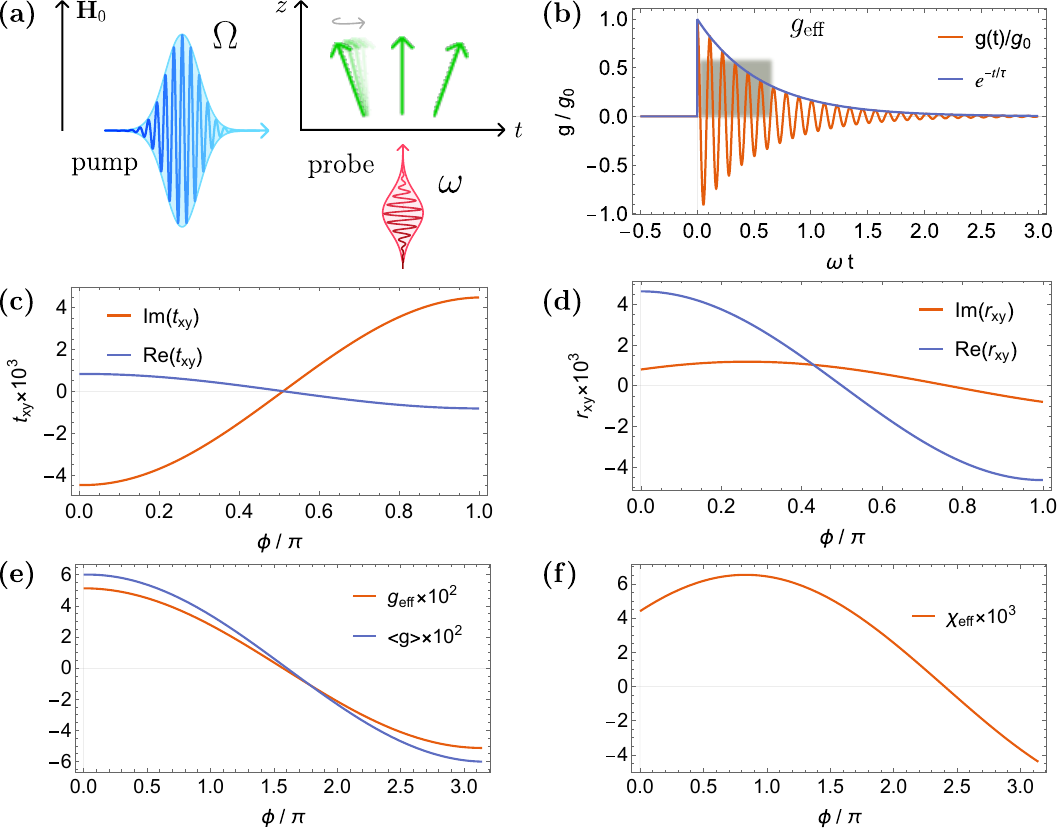}
  \caption{(a) Sketch of suggested geometry. Powerful high-frequency pump beam induces spin oscillations along the $z$ axis. Low-frequency probe signal propagates along $z$ experiencing time-reflection and time-transmission.
  (b) Decaying oscillations of gyrotropy approximated by the model of the time slab (gray rectangle) with the average gyrotropy $g_{\text{eff}}$ and axion response $\chi_{\text{eff}}$.
  (c,d) Simulated cross-polarized transmission and reflection of a probe wave versus the phase $\varphi$ of magnetization modulation.
  (e) Calculated average gyrotropy of the system $\left<g \right>$ and the effective gyrotropy $g_{\text{eff}}$ retrieved from transmission and reflection data versus the phase $\varphi$ of magnetization modulation.
  (f) Effective axion field retrieved from cross-polarized reflection versus the phase of gyrotropy oscillations.
  Simulation parameters: $\omega=2\pi$, $\Omega=0.09\;\omega$, $\tau=0.45$, $g_0=0.1$, $\eps=\mu=1$.}
  \label{Fig:Simulation}
\end{figure*}


To illustrate the applications of the developed concept, we consider an experimentally relevant situation when an intense high-frequency pulse excites the medium at the moment $t=0$ causing  rapid oscillations of magnetization with the frequency $\Omega$ and decay time $\tau$ [Fig.~\ref{Fig:Simulation}(a)]. This renders the material parameters of the medium time-dependent with  the gyrotropy
\begin{equation}\label{eq:Oscillations}
    g(t)=g_0\,e^{-t/\tau}\,\sin\left(\Omega t+\varphi\right)
\end{equation}
as illustrated in Fig.~\ref{Fig:Simulation}(b).


We assume that the traditional spectroscopic techniques do not allow to resolve those oscillations, so that neither average magnetization $\left<g\right>$ nor the initial phase of the oscillations $\varphi$ are known. Therefore, we explore an alternative route when throughout the interaction with the pump pulse the medium is also irradiated by the low-frequency probe signal. In that case, temporal modulation of the medium by the pump causes time transmission and time reflection of the probe. 

To illustrate that physics, we solve Eq.~\eqref{eq:WaveEq} numerically calculating cross-polarized transmission and reflection presented in  Fig.~\ref{Fig:Simulation}(c,d). Both quantities exhibit a profound dependence on the phase of magnetization oscillations. On the other hand, these optical characteristics can be retrieved experimentally.

Even though the oscillations of magnetization cannot be resolved directly, we demonstrate below that the measured cross-polarized time-transmission and time-reflection carry important signatures of ultrafast magnetization dynamics. As we prove~\cite{Supplement},  cross-polarized time-transmitted signal quantifies the average magnetization of the medium being independent on the induced effective axion field. In contrast, cross-polarized time-reflected signal depends both on the average magnetization and effective axion field. This allows to separate ultrafast magnetization dynamics~-- the source of effective axion field~-- from the persistent background magnetization.

Given the values of $T_{xy}$ and $R_{xy}$, it is possible to retrieve both average magnetization of the medium and effective axion field as illustrated in Fig.~\ref{Fig:Simulation}(e,f). Notably, the retrieved average gyrotropy agrees with the result obtained by the integration of Eq.~\eqref{eq:Oscillations}. Hence, having the experimental values of $T_{xy}$ and $R_{xy}$ and making the minimal assumptions regarding the magnetization dynamics, one can recover the phase $\varphi$ of ultrafast magnetization oscillations without resolving them.

These results highlight that the measurements of cross-polarized time-transmission and time-reflection provide a sensitive probe of magnetization dynamics providing a tool to explore ultrafast magnetic phenomena which themselves are an active area of research~\cite{Kimel2010}.



\section*{Discussion}

We have shown that rapid oscillations of magnetization in time can have a profound and observable effect on electromagnetic properties. Even if the oscillations themselves cannot be resolved due to the short timescales, their signatures~-- such as average magnetization and the initial phase~-- can still be retrieved from the characteristics of a time-reflected and time-transmitted probe signal with much lower frequency. This physics realizes the concept of an emergent dynamic axion field.

These results could provide an additional tool to probe dynamic axion fields in condensed matter via the measurement of time-reflection and time-transmission coefficients.

On the other hand, our results can find applications in the area of ultrafast magnetism. In particular, some magnetic structures feature spin precession frequency in the terahertz range~\cite{Kimel2018,Mikhaylovskiy2023} reaching 20~THz for KNiF$_3$~\cite{Bossini2016}. Due to high magnon frequencies, it is challenging to resolve these processes by the standard techniques. Our approach enables access to rapid magnetization changes using probe signals at a significantly lower frequency. A similar strategy could be useful to map structural phase transitions accompanied by the switching of magnetization.



\section*{Acknowledgments}

We acknowledge Prof. Alexandra Kalashnikova and Mr. Maxim Mazanov for valuable discussions and ITMO Center for Science Communication for help with illustrations. Theoretical models for effective axion response were supported by Priority 2030 Federal Academic Leadership Program. Numerical simulations were supported by the Russian Science Foundation, grant No.~23-72-10026. L.S. and M.A.G. acknowledge partial support from the Foundation for the Advancement of Theoretical Physics and Mathematics ``Basis''.

\bibliography{Timemod}

\end{document}


\title{Supplemental Materials: \\ Probing ultrafast magnetization dynamics via synthetic axion fields}

\author{Leon Shaposhnikov}
\email{These authors have equally contributed to this work}
\affiliation{School of Physics and Engineering, ITMO University, Saint  Petersburg 197101, Russia}

\author{Eduardo Barredo-Alamilla}
\email{These authors have equally contributed to this work}
\affiliation{School of Physics and Engineering, ITMO University, Saint  Petersburg 197101, Russia}

\author{Frank Wilczek}
\affiliation{Department of Physics, Stockholm University, Stockholm 10691, Sweden}
\affiliation{Center for Theoretical Physics, Massachusetts Institute of Technology, Cambridge, Massachusetts 02139, USA}
\affiliation{Department of Physics, Arizona State University, Tempe, Arizona 25287, USA}
\affiliation{Wilczek Quantum Center, Department of Physics and Astronomy, Shanghai Jiao Tong University, Shanghai 200240, China}

\author{Maxim A. Gorlach}
\email{m.gorlach@metalab.ifmo.ru}
\affiliation{School of Physics and Engineering, ITMO University, Saint  Petersburg 197101, Russia}


\maketitle
\widetext

\tableofcontents






\section{1. Floquet harmonics of the  fields and axion response}\label{app:Floquet}

In this section, we complement the derivation of the  axion response in the main text calculating Floquet harmonics of electric and magnetic fields associated with the rapid modulation of the medium magnetization. The medium is described by the material parameters
\begin{align}
\hat{\mu} & =
\begin{pmatrix}
\mu& -ig' & 0 \\
 ig' & \mu & 0 \\
 0 & 0 & \mu \\
\end{pmatrix}, &\hat{\eps} & = \eps.
\label{eq: mat par}
\end{align}
Here, $g'$ characterizes gyrotropic response and depends on the external magnetic field and intrinsic properties of the medium. In our analysis, we consider time-dependent gyrotropy $g'=g'(t)$ which is a periodic function of time $g'(t+T)=g'(t)$ with zero average, while $\eps$ and $\mu$ are assumed constant. 

In most physical situations the off-diagonal part of $\hat{\mu}$ tensor is small compared to the diagonal entries: $|g'|\ll\mu$. Thus, we may simplify the expression for $\hat{\mu}^{-1}$
\begin{equation}
   \hat{\mu}^{-1}\equiv \hat{\z}=
\left(
\begin{array}{ccc}
 \mu^{-1} & i g(t) & 0 \\
 -i g(t) & \mu^{-1} & 0 \\
 0 & 0 & \mu^{-1} \\
\end{array}
\right)+O(\frac{g'^2}{\mu^2}).
\label{InversePermeabilityApprox}\end{equation}
Here $g=g'/\mu^2$ is the effective gyrotropy of the medium which is used in the main text.
Next we expand $\hat{\zeta}$ into Fourier series where the Fourier harmonics with $m\neq0$ read:
\begin{align}\label{sup:eq: exact expression for z_m}
\hat{\z}_{m}=ig_m\cdot
\begin{pmatrix}
0 & 1 & 0\\
-1 & 0 & 0\\
0 & 0 & 0
\end{pmatrix}
=-i\,g_m\,\vc{e}_z^{\times}\:,
 \quad
g_m=\frac{1}{T}\int_0^Te^{i\Omega mt}g(t)dt
\end{align}
Here $\vc{e}_z^{\times}$ is matrix acting on a vector as a cross product by vector $\vc{e}_z$. As discussed in the main text, the equations capturing the physics of time-modulated medium take the following form:
\begin{equation}\label{eq:sup: wave div eq}
\begin{aligned}
&\frac{\eps}{c^2}\frac{\partial^2 \vc{B}}{\partial^2 t}+\rot\rot\vc{H}=0,\\
&
\divv\vc{B}=0,\\
&
{\bf H}=\hat{\mu}^{-1}\,{\bf B}\:.
\end{aligned}
\end{equation}
Since the parameters of the system are modulated in time with the period $T$, we seek the solution to  these equations in the form of the Floquet series:  $\vc{A}(\vc{r},t)=e^{i(\vc{kr}-\o t)}\sum\vc{A}_me^{-im\Omega t}$, where $\vc{A}$ is one of the fields  $\vc{E}\,,\vc{B}\,,\vc{D}\,,\vc{H}$, and $\Omega=2\pi/T$.

Since the medium has an axis of rotational symmetry, one of the components of the wave vector can always be set to zero. Thus, without loss of generality we consider $\vc{k}=(k_x,0,k_z)$. Equation~\eqref{eq:sup: wave div eq} then yields the set of linear equations for the respective Floquet harmonics
\begin{align}
&\label{eq: wave eq Bmx and Hmz Hmx} -k_zk_xH_{mz}+k_z^2H_{mx}=\eps q_m^2B_{mx},\\
&\label{eq: wave eq Bmy and Hmy}(k_x^2+k_z^2)H_{my}=\eps q_m^2B_{my},\\
&\label{eq: wave eq Bmz and Hmz Hmx} k_x^2H_{mz}-k_zk_xH_{mx}=\eps q_m^2B_{mz},\\
&\label{eq: div B Floquet}k_xB_{mx}+k_zB_{mz}=0\:,
\end{align}
the latter equation is also valid for $m=0$. Here $q=\o/c$ is the normalized frequency of the wave in the free space, while $q_m=q+Qm$ is the  normalized frequency of the respective Floquet harmonic, with $Q=\Omega/c$. We investigate the limit of rapid modulation, and hence $\xi=k/Q\ll 1$ plays role of the small parameter, while $q/Q$ and $k/Q$ have the same order of magnitude. 

Equations above together with the material equation ${\bf H}=\hat{\mu}^{-1}\,{\bf B}$ allow us to  derive the expressions for the Floquet harmonics including the first non-vanishing power of $\xi$. The material equation ${\bf H}=\hat{\mu}^{-1}\,{\bf B}$ yields the following relations for the Floquet harmonics:
\begin{align}
&  \label{eq: mat eq Hmx and Bmy Bmx} H_{mx}=\mu^{-1} B_{mx}+i\sum_{m'\neq m}g_{m-m'}B_{m'y},\\
& \label{eq: mat eq Hmy and Bmy Bmx}
H_{my}=\mu^{-1} B_{my}-i\sum_{m'\neq m}g_{m-m'}B_{m'x}, \\
&\label{eq: mat eq Hmz and Bmz} H_{mz}=\mu^{-1} B_{mz}.
\end{align}

From Eq.~\eqref{eq: wave eq Bmy and Hmy} we express: $B_{my}=k^2/\skob{Q^2m^2\eps}H_{my}+O(\xi^3)$, in the first non-zero order with respect to $\xi$, with $k^2=k_x^2+k_z^2$. Thus, Eq.~\eqref{eq: mat eq Hmy and Bmy Bmx} reads: $H_{my}=-ig_mB_{0x}+O(\xi^2)$, where $B_{0x}$ is the average of the $x$ component of the magnetic field ${\bf B}$. Combining the two equations above, we recover
\begin{equation}\label{eq: Bmy in B0x}
    B_{my}=-\frac{k^2}{m^2Q^2\eps}ig_mB_{0x}+O(\xi^3).
\end{equation}

Combining Eqs.~\eqref{eq: wave eq Bmx and Hmz Hmx},\eqref{eq: div B Floquet} and \eqref{eq: mat eq Hmz and Bmz}, we recover: $B_{mx}=k_z^2/\skob{m^2Q^2\eps}H_{mx}+O(\xi^3)$. As a result Eq.~\eqref{eq: mat eq Hmx and Bmy Bmx} reads: $H_{mx}=ig_mB_{0y}+O(\xi^2)$. From these two equations we derive another component of magnetic field Floquet harmonic:
\begin{equation}\label{eq: Bmx in B0y}
    B_{mx}=\frac{k_z^2}{m^2Q^2\eps}ig_mB_{0y}+O(\xi^3).
\end{equation}

Finally, making use of Eq.~\eqref{eq: div B Floquet} and Eq.~\eqref{eq: Bmx in B0y}, we derive the $z$-component of the magnetic field:
\begin{equation}\label{eq: Bmz in B0y}
    B_{mz}=-\frac{k_zk_x}{m^2Q^2\eps}ig_mB_{0y}+O(\xi^3).
\end{equation}

We can also explicitly write Floquet harmonics of the field ${\bf H}$:
\begin{gather}
    H_{mx}=ig_m\,B_{0y}+O(\xi^2)\:,\\
    H_{my}=-ig_m\,B_{0x}+O(\xi^2)\:,\\
    H_{mz}=-ig_m\,\frac{k_x\,k_z}{\eps\,\mu\,Q^2\,m^2}\,B_{0y}+O(\xi^3)\:.
\end{gather}

Next, using the expression $\rot{\bf H}=\frac{1}{c}\,\df{{\bf D}}{t}$, we compute the Fourier harmonics of electric displacement:
\begin{gather}
D_{mx}=-ig_m\,\frac{k_z}{mQ}\,B_{0x}+O(\xi^2)\:,\\
D_{my}=-ig_m\,\frac{k_z}{mQ}\,B_{0y}+O(\xi^2)\:,\\
D_{mz}=-ig_m\,\frac{k_z}{mQ}\,B_{0z}+O(\xi^2)\:,
\end{gather}
%
which can be recast in the vector form
\begin{equation}\label{eq: Dm in terms of B0}
    \vc{D}_m=-i\frac{g_m}{m}\frac{k_z}{Q}\vc{B}_0.
\end{equation}
Having the explicit expressions for the Floquet harmonics, we consider a time boundary keeping only the terms proportional to the first power of small parameter $\xi=k/Q$. Since all Floquet harmonics of magnetic field are proportional to $\xi^2$, the condition for magnetic field takes the conventional form
\begin{equation}\label{eq:BCondition}
    \left.{\bf B}\right|_{t=-0}={\bf B}_0\:.
\end{equation}
However, the condition for the averaged displacement is modified because of the contribution of nonzero higher-order harmonics $\vc{D}_m$:
\begin{equation}\label{eq:DCondition}
    \left.{\bf D}\right|_{t=-0}={\bf D}_0+\chi\,{\bf B}_0\:,
\end{equation}
where 
\begin{equation}\label{eq:FullAxion}
    \chi=-i\frac{k_z}{Q}\,\sum\limits_{m\not=0}\,\frac{g_m}{m}\:.
\end{equation}
As explained in the main text, the form of boundary condition Eq.~\eqref{eq:DCondition} allows to interpret $\chi$ as effective axion response with the magnitude defined by Eq.~\eqref{eq:FullAxion}.

We observe that $\chi$ depends on the direction of the wave propagation being maximal for $\vc{k}$ parallel to the magnetization direction (Faraday geometry) and vanishing when $\vc{k}$ is orthogonal to magnetization (Cotton-Mouton geometry), which resembles the physics of the conventional magneto-optical phenomena. Such dependence of $\chi$ on the direction of wave propagation is a manifestation of spatial dispersion effects intrinsic to time-modulated metamaterials. For clarity, from now on we focus on the situation when the wave propagates parallel to the direction of magnetization.




\section{2. Analysis of the effective axion response}

In this section, we recast the expression for the effective axion response Eq.~\eqref{eq:FullAxion} in a form more suitable for further analysis.
%
%
To this end, we transform the sum of the Fourier coefficients as
\begin{gather}
i\,\sum\limits_{m\not=0}\,\frac{g_m}{m}=i\,\sum\limits_{m\not=0}\,\frac{1}{m\,T}\,\int\limits_{0}^{T}\,g(t)\,e^{im\Omega t}\,dt=\frac{i}{T}\,\int\limits_{0}^{T}\,g(t)\,\left(\sum\limits_{m\not=0}\,\frac{e^{im\Omega t}}{m}\right)\,dt\notag\\
=
\frac{i}{T}\,\int\limits_{0}^{T}\,g(t)\,\left[\ln\left(1-e^{-i\Omega t}\right)-\ln\left(1-e^{i\Omega t}\right)\right]\,dt=-\frac{1}{T}\,\int\limits_0^T\,g(t)\,(\pi-\Omega \,t)\,dt=\frac{\Omega}{T}\,\int\limits_0^T\,g(t)\,t\,dt\:.
\end{gather}
Hence, for the case of periodic gyrotropy modulation the effective axion response can be recast in the form:
\begin{equation}\label{eq:AxionInt}
\chi=-\frac{\xi\,\Omega}{T}\,\int\limits_{0}^{T}\,g(t)\,t\,dt\:,
\end{equation}
where $\xi=k_z/Q$ and $Q=\Omega/c$. If the amplitude of $g(t)$ oscillations gradually changes in time, this expression can be generalized as
\begin{equation}
\chi=-\frac{\xi\,\Omega}{T}\,\int\limits_{t_0}^{t_0+T}\,g(t)\,\left(t-t_0\right)\,dt\:.
\end{equation}

Having the explicit expression for the effective axion field, we now calculate the magnitude of $\chi$ for several representative cases.

The first scenario analyzed in the main text assumes stepwise modulation of gyrotropy, i.e. $g(t)=g_0$ for $0<t<T/2$ and $g=-g_0$ for $T/2<t<T$, and then the pattern is repeated periodically. Straightforward calculation yields:
\begin{equation}
    \chi=g_0\,\xi\,\frac{\pi}{2}\:.
\label{sup4:effectiveaxionresponse}
\end{equation}
Note that depending on the sign of $g_0$, i.e. initial phase of the modulation, the effective axion response changes sign.


Next, we explore more physical situation when gyrotropy undergoes harmonic modulation over time according to the law $g(t)=g_0 \sin\left(\Omega t+\varphi\right)$. In this case, the effective axion response is given by:
%
%
\begin{equation}
    \chi=\xi\,g_0\,\cos\varphi\:.
\end{equation}

Interestingly, we observe the dependence of the effective axion field on the phase of gyrotropy modulation. We exploit this subtle effect to recover the phase $\varphi$ of the ultrafast magnetization oscillations.

\section{3. Effective permeability of axion metamaterial}

Time-varying gyrotropy results not only in the effective axion response, but also in the renormalization of the effective permeability of a  metamaterial proportional to $\xi^2$. For clarity, we focus here on the geometry of normal incidence (${\bf k}=k_z\,\bas$). To calculate that correction, we present the average ${\bf H}$ as
\begin{equation}\label{sup:eq: material eq in Metam up to n 2}
    \vc{H}_0=\sum_{m=-\infty}^\infty \hat{\zeta}_{-m}\vc{B}_{m}=\hat{\zeta}_0\vc{B}_0+\sum_{m\neq0} \hat{\zeta}_{-m}\vc{B}_{m}\:.
\end{equation}
As derived in Sec.~I, $\hat{\zeta}_0=\mu^{-1}$ and $\hat{\zeta}_m=-ig_m\,\bas^{\times}$ for $m\not=0$, Eq.~\eqref{sup:eq: exact expression for z_m}. In turn, Floquet harmonics ${\bf B}_m$ are given by Eqs.~\eqref{eq: Bmy in B0x},\eqref{eq: Bmx in B0y} and can be combined in the vector form as
\begin{equation}
        {\bf B}_m=-\frac{i\xi^2}{\eps\,m^2}\,g_m\,\left[\bas\times{\bf B}_0\right]+O(\xi^4)\:.
\end{equation}
This yields
\begin{equation*}  
\vc{H}_0=\skob{\mu^{-1}+\frac{\xi^2}{\eps}\,\sum\limits_{m\not=0}\frac{g_m\,g_{-m}}{m^2}\,\left[\hat{I}-\bas\times\bas\right]}\vc{B}_0+O(\xi^4).
\end{equation*}

Inverting this expression, we recover the effective permeability $\hat{\mu}_{\text{eff}}$ of the metamaterial:
\begin{align}
    \hat{\mu}_{\text{eff}} & = \mu\begin{pmatrix}
1-\alpha&0  &0  \\
0 &  1-\alpha&  0\\
 0&0  &1  \\
\end{pmatrix}, & \alpha & =\skob{\sum_{m'\neq0} \frac{g_mg_{-m}}{m^2}}\frac{\mu\,\xi^2}{\eps}.
\label{effectivepermeability1}
\end{align}
In the case of a step-like distribution $g(t)=g_0$ for $0<t<T/2$ and $g(t)=-g_0$ for $T/2<t<T$, one can show that 
\begin{equation}
    \alpha=\frac{\pi^2}{12}\,g_0^2\frac{\mu}{\eps}\xi^2.
\label{effectivepermeability2}\end{equation}

\section{4. Reflection and transmission at a single time boundary}

In this section, we derive the transmission and reflection matrices for a temporal boundary between two different axion media  with $\chi_1$, $\varepsilon_1$, $\mu_1$ for $t<t_0$ and $\chi_2$, $\varepsilon_2$, $\mu_2$  for $t>t_0$, respectively, as shown in Fig.~\ref{fig:axiontimeboundary}. Physically, this corresponds to the situation when the properties of the medium including its effective axion response, suddenly change modifying the propagation of plane waves.

The boundary conditions at a time boundary, Eqs.~(4) and (5) in the main text, are expressed as: 

\begin{figure}[h!]
  \centering
\includegraphics[width=0.6\textwidth]{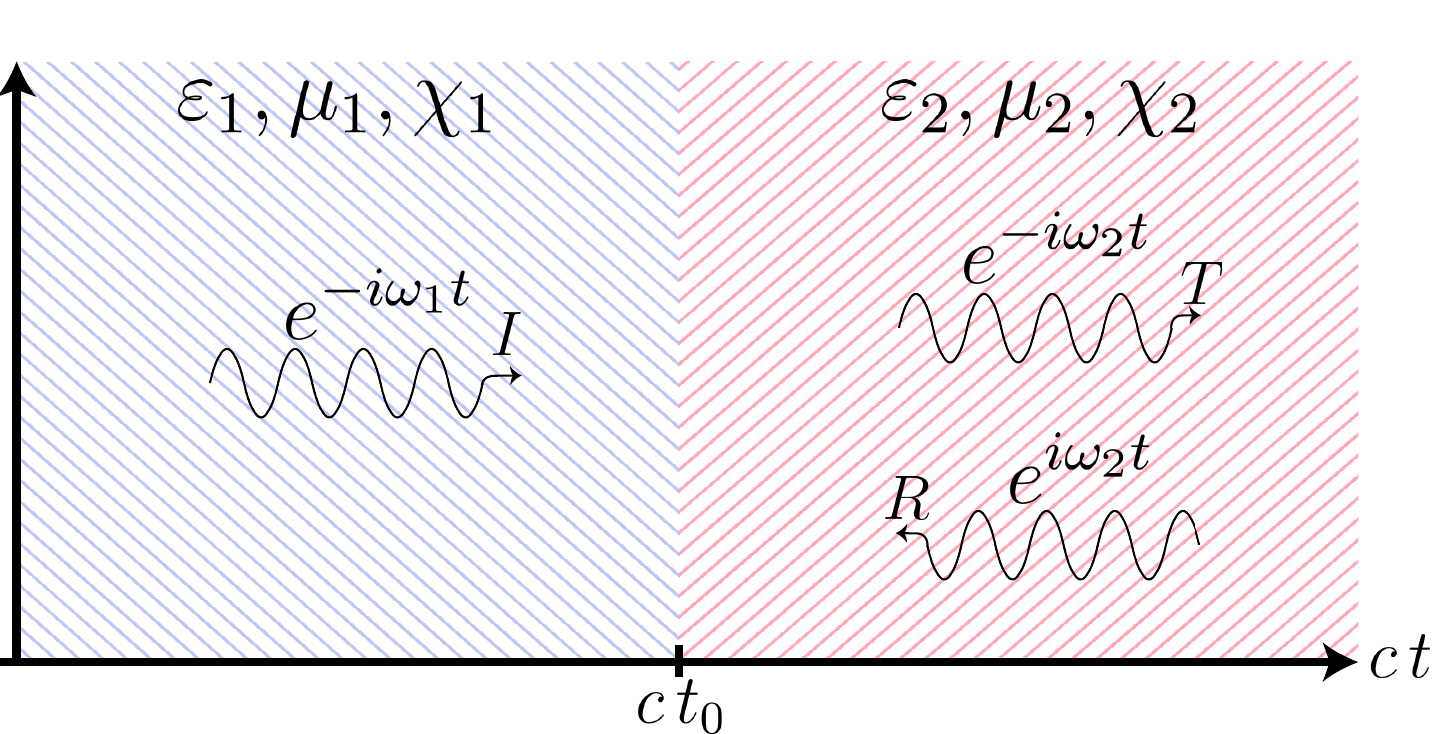}
  \caption{Reflection and transmission of a plane electromagnetic wave at a temporal boundary separating two different axion media.} 
  \label{fig:axiontimeboundary}
\end{figure}

\begin{align}
&\vc{B}^2=\vc{B}^1,\label{eq:BC in Temp boundary for B1}\\
    &\vc{D}^2-\vc{D}^1=\chi \vc{B}|_{t=t_0}, \label{eq:BC in Temp boundary for D1}
\end{align}
where $\chi = \chi_1-\chi_2$ and the conditions hold in all points of space. We consider plane waves propagating in the $\hat{\mathbf{z}}$ direction with the following representation:

\begin{align}
    \mathbf{B}^1 = & \mathbf{B}_{i}(z,t) = \mathbf{B}_i(k_i,\omega_i) e^{i k_i z} e^{-i \omega_i t}, \\
    \mathbf{B}^2 = & \mathbf{B}_{t}(z,t) + \mathbf{B}_{r}(z,t)= \mathbf{B}_t(k_t,\omega_t) e^{i k_t z} e^{-i \omega_t t} + \mathbf{B}_r(k_r,\omega_r) e^{i k_r z} e^{i \omega_r t},
\end{align}
where we use the subscripts $i$, $r$ and $t$ to tag the incident, reflected and transmitted wave, respectively. According to the time-boundary conditions in Eqs.~\eqref{eq:BC in Temp boundary for B1} and \eqref{eq:BC in Temp boundary for D1}, it is evident that $k_i = k_t = k_r=k$, $\omega_i = \omega_1 = ck/ \sqrt{\mu_1\varepsilon_1}$, and $\omega_r = \omega_t = \omega_2 = ck/ \sqrt{\mu_2\varepsilon_2}$. To find the amplitudes of reflected and transmitted fields, we use the boundary conditions which yield
%
%
\begin{align}
    \vc{B}_t + \vc{B}_r & = \vc{B}_i,\label{AxionTimeBoundary1}\\
    \sqrt{\frac{\varepsilon_2}{\mu_2}}(\vc{B}_t - \vc{B}_r) & = \sqrt{\frac{\varepsilon_1}{\mu_1}}\vc{B}_i + \chi \hat{\mathbf{n}} \times \mathbf{B}_i.\label{AxionTimeBoundary2}
\end{align}
Therefore, we can construct the time-boundary transfer matrix between two different media with axion responses $\chi_1$ and $\chi_2$, from Eqs. (\ref{AxionTimeBoundary1}),(\ref{AxionTimeBoundary2}) as

\begin{equation}
    \begin{pmatrix}
        \vphantom{\sqrt{\dfrac{\varepsilon_1}{\mu_1}}} \mathbf{B}_t +\mathbf{B}_r \\ \sqrt{\dfrac{\varepsilon_2}{\mu_2}} \Big(\mathbf{B}_t -\mathbf{B}_r\Big)  
    \end{pmatrix} =  \begin{pmatrix}
        \vphantom{\sqrt{\dfrac{\varepsilon_1}{\mu_1}}}M_{11} & M_{12}  \\
        M_{21} & \vphantom{\sqrt{\dfrac{\varepsilon_1}{\mu_1}}}M_{22}
    \end{pmatrix}
     \begin{pmatrix}
        \vphantom{\sqrt{\dfrac{\varepsilon_1}{\mu_1}}} \mathbf{B}_i \\ \sqrt{\dfrac{\varepsilon_1}{\mu_1}} \mathbf{B}_i
    \end{pmatrix}
\label{timeboundaryaxion12},\end{equation}
where $M_{11}=\mathbb{1}$, $M_{12}=0$, $M_{22}=\mathbb{1}$, and

\begin{equation}
    M_{21} = \chi \begin{pmatrix}
        \vphantom{\sqrt{\frac{\varepsilon_1}{\mu_1}}}0 & -1  \\
        1 & \vphantom{\sqrt{\frac{\varepsilon_1}{\mu_1}}}\hphantom{-}0
    \end{pmatrix}.
\end{equation}
Using the transfer-matrix between these two different media, we can compute the transmission and reflection matrices
\begin{align}
    \hat{T} & = \frac{1}{2\sqrt{\dfrac{\varepsilon_2}{\mu_2}}} \left[\sqrt{\frac{\varepsilon_2}{\mu_2}} M_{11} + \sqrt{\frac{\varepsilon_1\varepsilon_2}{\mu_1\mu_2}} M_{12}+ M_{21} + \sqrt{\frac{\varepsilon_1}{\mu_1}} M_{22}\right], \\
    \hat{R} & = \left[M_{11} +\sqrt{\frac{\varepsilon_1}{\mu_1}}M_{12} - \hat{T}\right],
\end{align}
that describe the amplitudes $\mathbf{B}_t = \hat{T}\;\mathbf{B}_i$ and $\mathbf{B}_r = \hat{R} \; \mathbf{B}_i$. The full form of these matrices is

\begin{align}
    \hat{T} & =  \frac{1}{2\sqrt{\dfrac{\varepsilon_2}{\mu_2}}} \begin{pmatrix}
 \sqrt{\dfrac{\varepsilon_2}{\mu_2}} +\sqrt{\dfrac{\varepsilon_1}{\mu_1}} & -\chi  \\
\chi & \sqrt{\dfrac{\varepsilon_2}{\mu_2}} +\sqrt{\dfrac{\varepsilon_1}{\mu_1}} \\
\end{pmatrix}, \\
    \hat{R} & = \frac{1}{2\sqrt{\dfrac{\varepsilon_2}{\mu_2}}}\begin{pmatrix}
 \sqrt{\dfrac{\varepsilon_2}{\mu_2}} -\sqrt{\dfrac{\varepsilon_1}{\mu_1}}& \chi  \\
-\chi & \sqrt{\dfrac{\varepsilon_2}{\mu_2}} -\sqrt{\dfrac{\varepsilon_1}{\mu_1}} \\
\end{pmatrix}.
\end{align}

Importantly, the jump in the effective axion response results in the emergence of cross-polarized components of reflected and transmitted waves, while the jump in permittivity and permeability is manifested in the co-polarized reflection and transmission.

\section{5. Reflection and transmission for an axion time slab}

\begin{figure}[h!]
  \centering
\includegraphics[width=0.6\textwidth]{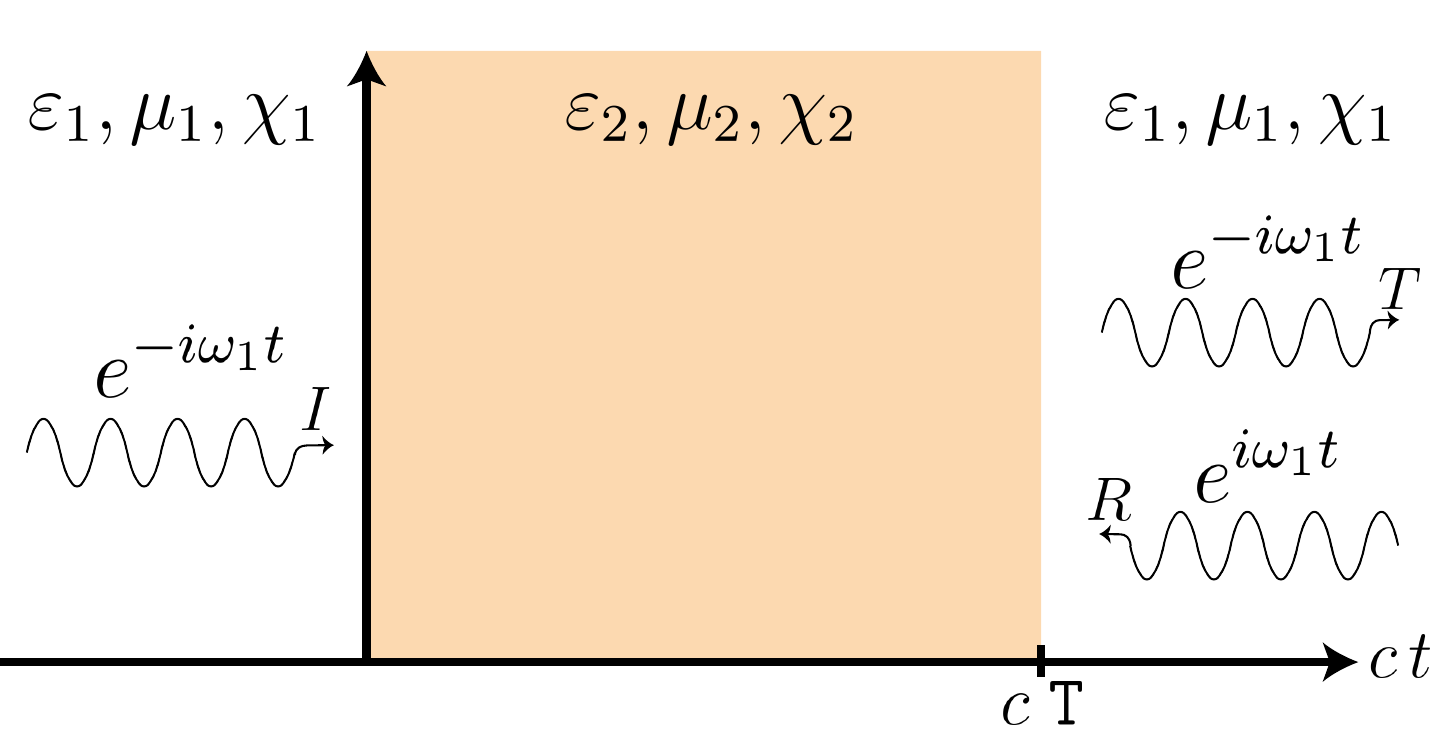}
  \caption{Reflection and transmission of a plane electromagnetic wave from a time slab with axion response} 
  \label{fig:axiontimeslab}
\end{figure}

Now we generalize the results of the previous section studying reflection and transmission of waves in the slab geometry. Physically this means that the properties of the medium are abruptly changed at a moment of time $t=0$ and then abruptly returned to their original values at the moment $t=\texttt{T}$. This situation is depicted schematically in Fig.~\ref{fig:axiontimeslab}.

The bulk propagation inside a pure-time axion slab of thickness $\texttt{T}$, can be written in the form
\begin{equation}
    \begin{pmatrix}
        \vphantom{\sqrt{\dfrac{\varepsilon_1}{\mu_1}}} \mathbf{B}_t(\texttt{T}) +\mathbf{B}_r(\texttt{T}) \\ \sqrt{\dfrac{\varepsilon}{\mu}} \Big(\mathbf{B}_t(\texttt{T}) -\mathbf{B}_r(\texttt{T})\Big)  
    \end{pmatrix} =  \begin{pmatrix}
        \vphantom{\sqrt{\dfrac{\varepsilon_1}{\mu_1}}} \cos(\omega' \texttt{T}) & -i \sin(\omega' \texttt{T}) \sqrt{\dfrac{\mu'}{\varepsilon'}} \\
        -i \sin(\omega' \texttt{T}) \sqrt{\dfrac{\varepsilon'}{\mu'}} & \cos(\omega' \texttt{T})
    \end{pmatrix}
     \begin{pmatrix}
        \vphantom{\sqrt{\dfrac{\varepsilon_1}{\mu_1}}} \mathbf{B}_i(0) \\ \sqrt{\dfrac{\varepsilon'}{\mu'}} \mathbf{B}_i(0)
    \end{pmatrix}
\label{timebulkaxion12}\:,\end{equation}
with the matrix in this equation capturing $e^{-i\omega t}$ phase advance of time-transmitted waves and $e^{i\omega t}$ time dependence of time-reflected waves.

Here, again, we exploit the transfer matrix  method. First, noticing Eq. (\ref{timeboundaryaxion12}), we write the transfer-matrix

\begin{equation}
    \hat{M} = \hat{M}_{2\rightarrow 1} \hat{M}(\texttt{T}) \hat{M}_{1\rightarrow 2} 
\end{equation}
with

\begin{align}
    \hat{M}_{1\rightarrow 2 } & =  \begin{pmatrix}
        \hphantom{-}1 & \hphantom{-}0 & \hphantom{-}0 & \hphantom{-}0 \\
        \hphantom{-}0 & \hphantom{-}1 & \hphantom{-}0 & \hphantom{-}0 \\
        \hphantom{-}0 & -\chi & \hphantom{-}1 & \hphantom{-}0 \\
        \hphantom{-}\chi & \hphantom{-}0 & \hphantom{-}0 & \hphantom{-}1
    \end{pmatrix}, & 
    \hat{M}_{2\rightarrow 1 } & = \begin{pmatrix}
        \hphantom{-}1 & \hphantom{-}0 & \hphantom{-}0 & \hphantom{-}0 \\
        \hphantom{-}0 & \hphantom{-}1 & \hphantom{-}0 & \hphantom{-}0 \\
        \hphantom{-}0 & \hphantom{-}\chi & \hphantom{-}1 & \hphantom{-}0 \\
        -\chi & \hphantom{-}0 & \hphantom{-}0 & \hphantom{-}1
    \end{pmatrix}
\label{timeboundaryaxiontwoways},\end{align}
and
\begin{equation}
    \hat{M}(\texttt{T}) =
    \begin{pmatrix} 
        \vphantom{\sqrt{\dfrac{\varepsilon_1}{\mu_1}}} \cos(\omega_2 \texttt{T}) & -i \sin(\omega_2 \texttt{T}) \sqrt{\dfrac{\mu_2}{\varepsilon_2}} \\
        -i \sin(\omega_2 \texttt{T}) \sqrt{\dfrac{\varepsilon_2}{\mu_2}} & \vphantom{\sqrt{\dfrac{\varepsilon_1}{\mu_1}}} \cos(\omega_2 \texttt{T})
    \end{pmatrix},
\end{equation}
where $\omega_2={kc}/{\sqrt{\mu_2\varepsilon_2}}$, and considering two flat time-boundaries at $t=0$ and $t = \texttt{T}$. The components of the reflection and transmission coefficients for the slab are

\begin{align}
    T_{xx} \hphantom{-} = \hphantom{-}T_{yy} & = \cos (\omega_2 \texttt{T}) - \frac{i}{2} \left(\sqrt{\dfrac{\mu_1 \varepsilon_2}{\mu_2\varepsilon_1}} + \sqrt{\dfrac{\mu_2\varepsilon_1}{\mu_1\varepsilon_2}} + \sqrt{\dfrac{\mu_1 \mu_2}{\varepsilon_1\varepsilon_2}}\chi^2 \right) \sin(\omega_2 \texttt{T}), \label{Txx:puretimeaxionslab}\\
    T_{xy} \hphantom{-} = -T_{yx} & = 0, \label{Txy:puretimeaxionslab}\\
    R_{xx} \hphantom{-} = \hphantom{-}R_{yy} & = \frac{i}{2} \left(\sqrt{\dfrac{\mu_1 \varepsilon_2}{\mu_2\varepsilon1}} - \sqrt{\dfrac{\mu_2\varepsilon_1}{\mu_1\varepsilon_2}} + \sqrt{\dfrac{\mu_1 \mu_2}{\varepsilon_1\varepsilon_2}}\chi^2 \right) \sin(\omega_2 \texttt{T}), \label{Rxx:puretimeaxionslab} \\
    R_{xy} \hphantom{-} = - R_{yx} & = -i \sqrt{\dfrac{\mu_2 }{\varepsilon_2}} \chi \sin(\omega_2 \texttt{T}).\label{Rxy:puretimeaxionslab}
\end{align}

Interestingly, the effective axion field is manifested in cross-polarized reflection. However, cross-polarized transmission is fully suppressed.

\section{6. Effective axion response of a time-modulated gyrotropic medium}

In this section, we present the results of numerical simulations for gyrotropic medium with the periodic stepwise modulation of gyrotropy. We analyze the metamaterial behavior using the transfer matrix method and compare the results with the effective model replacing time-varying gyrotropy by the constant effective axion field, Eq.~(\ref{sup4:effectiveaxionresponse}). We consider a gyrotropic material with the magnetization parallel to the $z$-axis and  periodically modulated in time: 

\begin{align}\label{eq: gyro mu mat par}
\hat{\mu} & = \begin{pmatrix}
 \mu & -i g(t) &  0\\
ig(t) & \mu &0  \\
 0&0  &\mu  \\
\end{pmatrix}, & \hat{\eps} & =\eps.
\end{align}
Here, $g(t)$ corresponds to $g'$ in designations of Eq.~\eqref{eq: mat par}.  We assume that $\varepsilon(t)$, $\mu(t)$ and $g(t)$ are real, periodic and non-dispersive. The modulation period is given by $\Delta \tau = \tau_A + \tau_B$, while the periodicity condition is expressed as
\begin{equation}
    \begin{pmatrix}
        \varepsilon(t) \\
        \mu(t) \\
        g(t) \\
    \end{pmatrix} = 
    \begin{pmatrix}
        \varepsilon(t + \Delta \tau) \\
        \mu(t + \Delta \tau) \\
        g(t + \Delta \tau)
    \end{pmatrix}.
\end{equation}
The photonic time crystal is made of two layers $A$ and $B$ that correspond to the different values of material parameters. Therefore the permeability tensor is
\begin{equation}
    \mu_\alpha^{ij} = 
    \begin{pmatrix}
        \mu_\alpha & -i g_\alpha &  0\\
        ig_\alpha & \mu_\alpha &0  \\
         0 & 0  &\mu_\alpha  \\
    \end{pmatrix},
\end{equation}
where $\alpha = A,B$. In this case we consider $g_B = -g_A = -g$ that corresponds to the stepwise modulation of magnetization with the zero average. The constitutive relations for a single layer $\alpha=A,B$, written in terms of $B^\alpha_i$, and taking $g\ll\mu$, we have

\begin{align}
    H^i_\alpha & = (\mu_\alpha^{ij})^{-1} B_\alpha^j, &  (\mu_\alpha^{ij})^{-1} \approx & \frac{1}{(\mu_\alpha)^2} 
    \begin{pmatrix}
        \mu_\alpha & i g_\alpha &  0\\
        -ig_\alpha & \mu_\alpha &0  \\
         0 & 0  &\mu_\alpha  \\
    \end{pmatrix}
\end{align}

Using a chiral basis, $\mathbf{B}_\alpha = B_\alpha \,\hat{\vc{x}}+ i\eta B_\alpha \,\hat{\vc{y}}$, where $\eta = \pm 1$ for right- and left-handed polarization, we can write the $H$ fields as

\begin{align}
    H_{\alpha,\eta} & = (\mu_{\alpha,\eta})^{-1} B_{\alpha,\eta}, & (\mu_{\alpha,\eta})^{-1} =  \frac{1}{\mu^2_\alpha} (\mu_\alpha - \eta g_\alpha).
\end{align}
Next, we construct the magnetic field propagating in a layer $\alpha=A,B$ as~\cite{Araujo2021Nov}

\begin{equation}
    \mathbf{B}_{\alpha,\eta} = B_{\alpha,\eta}\; e^{-i \omega_{\alpha,\eta} t} \hat{{v}}_\eta + B'_{\alpha,\eta} \; e^{i \omega_{\alpha,\eta} t} \hat{{v}}_\eta
\end{equation}
and
\begin{equation}
    \mathbf{D}_{\alpha,\eta} = i\eta \sqrt{\dfrac{\varepsilon_\alpha}{\mu_{\alpha,\eta}}} B_{\alpha,\eta}\; e^{-i \omega_{\alpha,\eta} t} \hat{{v}}_\eta - i\eta \sqrt{\dfrac{\varepsilon_\alpha}{\mu_{\alpha,\eta}}} B'_{\alpha,\eta}\; e^{i \omega_{\alpha,\eta} t} \hat{{v}}_\eta,
\end{equation}
where 
\begin{equation}
    \omega_{\alpha,\eta} = \frac{c k}{ \sqrt{\varepsilon_\alpha \mu_{\alpha, \eta}}},
\label{dispersionrelationgyrolayer}\end{equation}
is the frequency of the wave where $k$ is the propagation constant, $B_{\alpha,\eta}$ and $B'_{\alpha,\eta}$ are the amplitudes of the time-forward and time-backward waves, respectively. Here, $\hat{v}_\eta= \hat{v}_\pm$ is given by
\begin{equation}
    \hat{v}_\eta = \frac{1}{\sqrt{2}}
    \begin{pmatrix}
        1 \\
        i \eta \\
        0 
    \end{pmatrix}.
\end{equation}
We know from the conventional temporal-boundary conditions that $\mathbf{B}$ and $\mathbf{D}$ are continuous, therefore

\begin{align}
    B_A + B'_A & = B_B + B'_B,\label{boundarygyrotropictime1} \\
    \xi_A B_A -  \xi_A B'_A & =  \xi_B  B_B - \xi_B  B'_B, \label{boundarygyrotropictime2}
\end{align}
where $\xi_\alpha = i\eta \sqrt{\varepsilon_\alpha / \mu_{\alpha,\eta}}$. We will omit the dependence on the polarization, to compute the $2 \times 2$ transfer matrix for a temporal-boundary between a medium $A$ to $B$ using Eqs. (\ref{boundarygyrotropictime1}),(\ref{boundarygyrotropictime2}) 

\begin{align}
    \begin{pmatrix}
        B_B\\
        B'_B
    \end{pmatrix} & = M_{AB} 
    \begin{pmatrix}
        B_A\\
        B'_A
    \end{pmatrix}, & M_{AB} & = \frac{1}{2}
    \begin{pmatrix}
        1 + \dfrac{Y_A}{Y_B} &  1 - \dfrac{Y_A}{Y_B} \\
        1 - \dfrac{Y_A}{Y_B} &  1 + \dfrac{Y_A}{Y_B}
    \end{pmatrix}.
\end{align}
Next we consider a basis with the amplitude vector $(B_{\alpha,+},B'_{\alpha,+},B_{\alpha,-},B'_{\alpha,-})^{T}$, where $+(-)$ describes the right- (left-) handed polarization. The $4\times 4$ interface matrix is given by

\begin{equation}
    \begin{pmatrix}
         \vphantom{\dfrac{Y_{A,+}}{Y_{B,+}}} B_{B,+} \\
         \vphantom{\dfrac{Y_{A,+}}{Y_{B,+}}} B'_{B,+} \\
         \vphantom{\dfrac{Y_{A,+}}{Y_{B,+}}} B_{B,-} \\
         \vphantom{\dfrac{Y_{A,+}}{Y_{B,+}}} B'_{B,-}
    \end{pmatrix} = \frac{1}{2}
    \begin{pmatrix}
         1 + \dfrac{Y_{A,+}}{Y_{B,+}} &  1 - \dfrac{Y_{A,+}}{Y_{B,+}} & 0 & 0 \\
         1 - \dfrac{Y_{A,+}}{Y_{B,+}} & 1 + \dfrac{Y_{A,+}}{Y_{B,+}}&  0 & 0 \\
         0 & 0 & 1 + \dfrac{Y_{A,-}}{Y_{B,-}} &  1 - \dfrac{Y_{A,-}}{Y_{B,-}} \\
         0 & 0 & 1 - \dfrac{Y_{A,-}}{Y_{B,-}} &  1 + \dfrac{Y_{A,-}}{Y_{B,-}}
    \end{pmatrix}
    \begin{pmatrix}
         \vphantom{\dfrac{Y_{A,+}}{Y_{B,+}}} B_{A,+} \\
         \vphantom{\dfrac{Y_{A,+}}{Y_{B,+}}} B'_{A,+} \\
         \vphantom{\dfrac{Y_{A,+}}{Y_{B,+}}} B_{A,-} \\
         \vphantom{\dfrac{Y_{A,+}}{Y_{B,+}}} B'_{A,-}
    \end{pmatrix},
\end{equation}
where we define $Y_{\alpha, \eta} = \sqrt{\varepsilon_\alpha / \mu_{\alpha,\eta}}$. Finally, for the case of electromagnetic wave propagating inside the medium $\alpha=A,B$, with thickness $\tau_\alpha$ and frequency $\omega_{\alpha,\eta}$, the $4\times 4$ bulk-propagation matrix is given by

\begin{equation}
    M_\alpha = M_\alpha(\tau_\alpha) = 
    \begin{pmatrix}
        e^{-i \omega_{\alpha,+} \tau_\alpha} & 0 & 0 & 0 \\
        0 & e^{i \omega_{\alpha,+} \tau_\alpha} & 0 & 0 \\
        0 & 0 & e^{-i \omega_{\alpha,-} \tau_\alpha} & 0 \\
        0 & 0 & 0 & e^{i \omega_{\alpha,-} \tau_\alpha}
    \end{pmatrix}
\end{equation}
The resultant transfer matrix describing the temporal-multilayered  structure with even number of $2N$ layers then reads

\begin{equation}
    M_{\text{slab}} = M_{B-\text{air}} M_{AB} (M_{BA} M_{B} M_{AB} M_{A})^{2N} M_{\text{air}-A}  = 
    \begin{pmatrix}
        M^+_{\text{slab}} & 0 \\
        0 & M^-_{\text{slab}}
    \end{pmatrix}
\end{equation}

This matrix describes the relation between the incident, time-transmitted and time-reflected waves:
\begin{align}
    \begin{pmatrix}
        B^t_\eta \\
        B^r_\eta
    \end{pmatrix} & = M^\eta_{\text{slab}} 
    \begin{pmatrix}
        B^i_\eta \\
        0
    \end{pmatrix}, & M^\eta_{\text{slab}} & =
    \begin{pmatrix}
        M^\eta_{11} & M^\eta_{12} \\
        M^\eta_{21} & M^\eta_{22}
    \end{pmatrix}
\end{align}

Hence, we deduce the amplitude of the transmitted and reflected wave for both polarizations 

\begin{align}
    B^t_\eta & = M^\eta_{11} B^i_\eta &  B^r_\eta & = M^\eta_{21} B^i_\eta.
\end{align}

Assuming that the incident wave is polarized along $x$ axis, we write the transmitted and reflected amplitudes of fields $\mathbf{B}$ as

\begin{eqnarray}
    \mathbf{B}^t & = & \frac{1}{2}(M^+_{11} + M^-_{11}) B^i \, \hat{\vc{x}} + \frac{i}{2}(M^+_{11} - M^-_{11}) B^i \, \hat{\vc{y}} \\
    \mathbf{B}^r & = & \frac{1}{2}(M^+_{21} + M^-_{21}) B^i \, \hat{\vc{x}} + \frac{i}{2}(M^+_{21} - M^-_{21}) B^i \, \hat{\vc{y}}
\end{eqnarray}
which yields the co- and cross-polarized transmission and reflection in terms of the components of the transfer-matrix for the full multilayered slab
\begin{align}
    T_{xx} & = \frac{1}{2}\Bigg( M^+_{11} + M^-_{11}\Bigg), & T_{xy} & = \frac{1}{2i}\Bigg( M^+_{11} - M^-_{11}\Bigg), \label{T:photoniccrystal}\\
    R_{xx} & = \frac{1}{2}\Bigg( M^+_{21} + M^-_{21}\Bigg), & R_{xy} & = \frac{1}{2i}\Bigg( M^+_{21} - M^-_{21}\Bigg).\label{R:photoniccrystal}
\end{align}

\begin{figure}[t!]
\centering
\includegraphics[width=0.98 \textwidth]{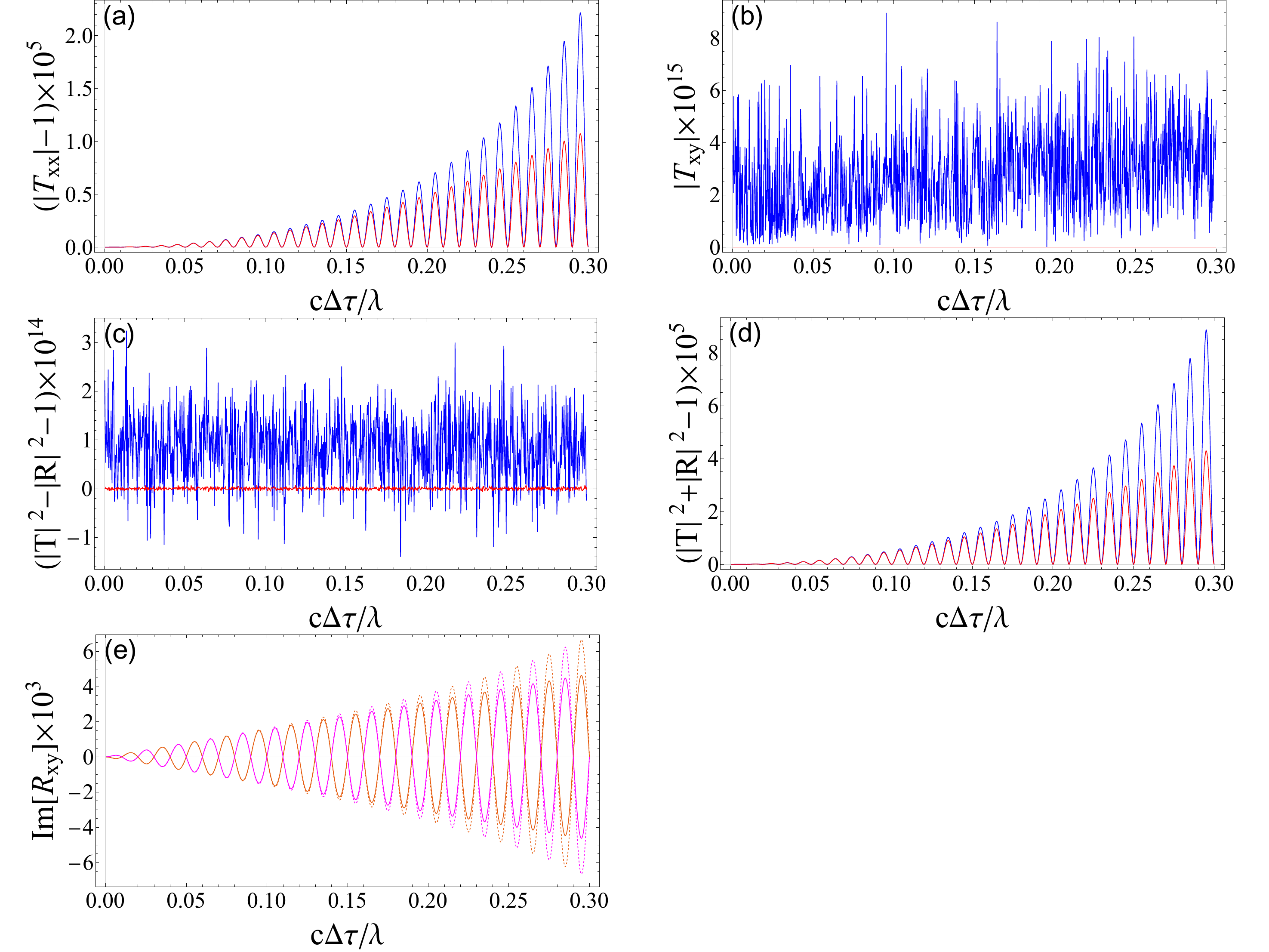}
\caption{Validation of the effective axion response. (a-d) Red lines correspond to the effective medium approximation and blue lines to the results of the transfer-matrix method applied to the multilayered structure. Panel (a) corresponds to the co-polarized transmission, (b) shows the cross-polarized transmission coefficient which is essentially zero, panel (c) illustrates the conservation of momentum $|T|^2-|R|^2 = 1$, and panel (d) suggests that energy is not conserved $|T|^2+|R|^2\neq 1$. Panel (e) shows the imaginary part of cross-polarized reflection, where orange and magenta lines correspond to the initial positive and negative gyrotropy, respectively, both with $|g|=0.01$. Solid lines correspond to the effective medium approximation, and the dashed lines to the real multilayered structure. Overall, 50 modulation periods are considered.}
\label{ValidationPureTimeAxionSlab}
\end{figure}

For numerical validation of the effective medium approach we consider photonic time crystal (PTC) consisting of 50 unit cells, i.e. 50 full modulation periods. Each unit cell includes two gyrotropic layers of equal thickness, $\Delta \tau/2$, with the opposite magnetization ($g = 0.01$). For simplicity, we assume that the permittivity and permeability of the gyrotropic layers are both equal to 1. The transmission and reflection coefficients for the photonic crystal are evaluated using Eqs~(\ref{T:photoniccrystal}),(\ref{R:photoniccrystal}).

On the other hand, the same quantities can be computed in the effective medium framework. Note that the effective axion response depends on the modulation's starting point [Eqs.~(\ref{sup4:effectiveaxionresponse}),(\ref{effectivepermeability1})]. Initiating 
the modulation from positive gyrotropy results in an effective positive axion response $\chi_\text{eff}$ and an effective permeability 
$\mu_\text{eff}$ given by:  
\begin{align}
    \chi_\text{eff} & =  \frac{\pi}{2} \, \frac{g_0}{ \mu_\text{eff}^2} \,\xi, & \mu_\text{eff} & =  \mu\left( 1-\frac{\pi^2}{12}|g_0|^2\frac{\mu}{\eps} \xi^2\right),
\label{effectiveaxionfieldandmu}\end{align}
where $g_0$ is the gyrotropy of the first layer, $\xi(\lambda) = k_z/Q = c\Delta \tau/\lambda$ is the period-to-wavelength ratio, and $\lambda$ is the vacuum wavelength. The permittivity remains unchanged, $\varepsilon_\text{eff} = \varepsilon$. The transmission and reflection coefficients for an axion slab have the analytical forms given by Eqs.~(\ref{Txx:puretimeaxionslab})-(\ref{Rxy:puretimeaxionslab}). For the permittivity and permeability, we set $\varepsilon_1 = \varepsilon_2 = 1$, $\mu_1 = 1$, and $\mu_2 = \mu_\text{eff}$ to model the pure-time axion slab in vacuum. The temporal path inside the slab is given by:

\begin{equation}
\omega_2 \texttt{T} = \frac{2\pi}{\sqrt{\mu_{\text{eff}} \, \varepsilon_{\text{eff}}}} N \xi,
\end{equation}
where $N = 50$ represents the number of periods in the multilayered structure.

The thickness of the slab is fixed at $\texttt{T} = N \Delta \tau$. By varying the wavelength $\lambda$, we can plot the reflection and transmission coefficients both for the photonic crystal and the effective axion medium in terms of the period-to-wavelength ratio $\xi$. In the main text [Fig. 2], we demonstrate the validation of the effective axion response. It is evident that for rapid modulation with $\xi \ll 1$, the transmission and reflection coefficients of the multilayered structure closely match those of the effective axion medium. As $\xi$ increases, the effective medium solution starts to deviate from the real multilayered structure. In Fig. \ref{ValidationPureTimeAxionSlab}(a-d) we present results complementing those in the main text like co- and cross-polarized transmission. We observe that the cross-polarized transmission is essentially zero. We also recover the conservation of momentum as $|T|^2-|R|^2 = 1$ derived from the fact that the wave-vector $k$ remains constant, and the non-conservation of energy $|T|^2+|R|^2 \neq 1$ that arises due to the temporal modulation of the medium. 

Finally, in Fig. \ref{ValidationPureTimeAxionSlab}(e) we verify the dependence of cross-polarized time-reflection on the gyrotropy of the first layer of photonic time crystal. Specifically, if the first layer has positive (negative) gyrotropy, the effective axion field is also positive (negative), see Eq.~(\ref{effectiveaxionfieldandmu}).

\section{7. Simulation of reflection and transmission for the time-varying effective axion field $\chi(t) = b_0 t$}


In this section, we examine the situation when the magnetization experiences rapid modulation with an amplitude linearly growing in time. From the effective medium perspective, this describes time-dependent effective axion field $\chi(t)=b_0 t$ that corresponds to a Weyl semimetal with energy separation of the Weyl nodes equal to $b_0$.

First, consider the bulk propagation inside a time-slab with the axion field linearly growing in time: $\chi(t) = b_0 t$. Using the equations of axion electrodynamics, we recover two propagating modes in the long-wavelength region ($kc > b_0 \mu$).

\begin{equation}
\tilde{\omega}_\eta = \sqrt{\frac{k^2c^2}{\mu\varepsilon} - \eta \frac{b_0 kc}{\varepsilon}} = \frac{kc}{\sqrt{\mu\varepsilon}} \sqrt{1-\eta \frac{\mu b_0}{kc}},
\end{equation}
is the frequency of the wave with the propagation constant $k$. The electromagnetic wave is circularly polarized with $\eta = 1$ and $\eta = -1$ corresponding to right- and left-handed modes. Henceforth, we can write the fields in terms of the circular polarization as

\begin{equation}
    \mathbf{B}_\eta = B_\eta e^{-i \tilde{\omega}_\eta t} \hat{v}_\eta +B_\eta' e^{i \tilde{\omega}_\eta t} \hat{v}_\eta,
\end{equation}
and

\begin{equation}
    \mathbf{D}_{\eta} = i\eta \sqrt{\dfrac{\varepsilon}{\mu}}\sqrt{1-\eta \frac{\mu b_0}{kc}} B_{\eta}\; e^{-i \tilde{\omega}_{\eta} t} \hat{{v}}_\eta - i\eta \sqrt{\dfrac{\varepsilon}{\mu}}\sqrt{1-\eta \frac{\mu b_0}{kc}} B_{\eta}\; e^{i \tilde{\omega}_{\eta} t} \hat{{v}}_\eta,
\end{equation}
where $B_{\eta}$ and $B'_{\eta}$ are the amplitudes of the forward and backward waves, respectively. Here, $\hat{v}_\eta= \hat{v}_\pm$ is given by
\begin{equation}
    \hat{v}_\eta = \frac{1}{\sqrt{2}}
    \begin{pmatrix}
        1 \\
        i \eta \\
        0 
    \end{pmatrix}
\end{equation}

Consider a temporal boundary between two different ponderable axion media  with $\chi_1(t) = b_{0,1} \, t$, $\varepsilon_1$, $\mu_1$ for $ct<ct_0$ and $\chi_2(t) = b_{0,2} \,t$, $\varepsilon_2$, $\mu_2$  for $ct>ct_0$, respectively. The boundary conditions Eqs. (\ref{eq:BC in Temp boundary for B1}),(\ref{eq:BC in Temp boundary for D1}) take the form
\begin{align}
    B_{1,\eta} + B'_{1,\eta} & = B_{2,\eta} + B'_{2,\eta}, \label{boundaryweyltime1} \\
    z_{1,\eta} B_{1,\eta} -  z_{1,\eta} B'_{1,\eta} & =  (z_{2,\eta} - \chi)  B_{2,\eta} - (z_{2,\eta} +\chi)  B'_{2,\eta}, \label{boundaryweyltime2}
\end{align}
where  $\chi(t_0) = \chi_1(t_0)-\chi_2(t_0)$, and 
\begin{equation}
z_{\alpha,\eta} = i\eta \sqrt{\dfrac{\varepsilon_\alpha}{\mu_\alpha}}\sqrt{1-\eta \frac{\mu_\alpha b_{0,\alpha}}{kc}}.
\end{equation}
This allows to construct $2 \times 2$ transfer matrix for a temporal-boundary of two different ponderable axion media:
\begin{align}
    \begin{pmatrix}
       \vphantom{\dfrac{z_1-\chi}{z_2}} B_{2,\eta}  \vspace{5pt}\\
        \vphantom{\dfrac{z_1-\chi}{z_2}}B'_{2,\eta}
    \end{pmatrix}  =\frac{1}{2} \begin{pmatrix}
        1 + \dfrac{z_{1,\eta}+\chi}{z_{2,\eta}} &  1 - \dfrac{z_{1,\eta}-\chi}{z_{2,\eta}} \vspace{5pt}\\
        1 - \dfrac{z_{1,\eta}+\chi}{z_{2,\eta}} &  1 + \dfrac{z_{1,\eta}-\chi}{z_{2,\eta}}
    \end{pmatrix} 
    \begin{pmatrix}
       \vphantom{\dfrac{z_1-\chi}{z_2}} B_{1,\eta}  \vspace{5pt} \\
        \vphantom{\dfrac{z_1-\chi}{z_2}} B'_{1,\eta}
    \end{pmatrix}. 
\end{align}
Finally, we write the full transfer matrix for light impinging a pure-time Weyl semimetal slab from vacuum as

\begin{align}
     M^\eta_{\text{WS-slab}} & = \hat{M}^\eta_{2\rightarrow 1} (t=\texttt{T})M^\eta_\text{Bulk}(\texttt{T}) \hat{M}^\eta_{1\rightarrow 2} (t=0), &  M_\text{Bulk}(\texttt{T}) = & 
    \begin{pmatrix}
        e^{-i \tilde{\omega}_\eta \texttt{T} } & 0 \\
        0 & e^{i \tilde{\omega}_\eta \texttt{T}} 
    \end{pmatrix}
\end{align}
where 

\begin{align}
    \hat{M}_{1\rightarrow 2 }(t=0) & =\frac{1}{2}\begin{pmatrix}
        1 + \dfrac{z_{1,\eta}}{z_{2,\eta}} &  1 - \dfrac{z_{1,\eta}}{z_{2,\eta}} \vspace{5pt}\\
        1 - \dfrac{z_{1,\eta}}{z_{2,\eta}} &  1 + \dfrac{z_{1,\eta}}{z_{2,\eta}}
    \end{pmatrix}, &  \hat{M}_{2\rightarrow 1 } & =\frac{1}{2}\begin{pmatrix}
        1 + \dfrac{z_{2,\eta}+b_0 \texttt{T}}{z_{1,\eta}} &  1 - \dfrac{z_{2,\eta}-b_0 \texttt{T}}{z_{1,\eta}} \vspace{5pt}\\
        1 - \dfrac{z_{2,\eta}+b_0 \texttt{T}}{z_{1,\eta}} &  1 + \dfrac{z_{2,\eta}-b_0 \texttt{T}}{z_{1,\eta}}
    \end{pmatrix},
\end{align}
with 
\begin{align}
    z_{1,\eta} & = i\eta \sqrt{\dfrac{\varepsilon_0}{\mu_0}}, &   z_{2,\eta} & = i\eta \sqrt{\dfrac{\varepsilon_0}{\mu_0}}\sqrt{1-\eta \frac{\mu_0 b_{0}}{kc}},
\end{align}
This matrix connects incident, time-reflected and time-transmitted fields as

\begin{align}
    \begin{pmatrix}
        B^t_\eta \\
        B^r_\eta
    \end{pmatrix} & =  M^\eta_{\text{WS-slab}}
    \begin{pmatrix}
        B^i_\eta \\
        0
    \end{pmatrix}, &  M^\eta_{\text{WS-slab}} & =
    \begin{pmatrix}
        M^\eta_{11} & M^\eta_{12} \\
        M^\eta_{21} & M^\eta_{22}
    \end{pmatrix}
\end{align}
Just as in the pure-time axion slab case, we can write the transmission and reflection coefficients as 
\begin{align}
    T_{xx} & = \frac{1}{2}\Bigg( M^+_{11} + M^-_{11}\Bigg), & T_{xy} & = \frac{1}{2i}\Bigg( M^+_{11} - M^-_{11}\Bigg), \label{T:photoniccrystal2}\\
    R_{xx} & = \frac{1}{2}\Bigg( M^+_{21} + M^-_{21}\Bigg), & R_{xy} & = \frac{1}{2i}\Bigg( M^+_{21} - M^-_{21}\Bigg).\label{R:photoniccrystal2}
\end{align}
by assuming that the initial magnetic field is polarized along the $x$ axis.

\begin{figure}[b!]
  \centering
\includegraphics[width=0.90\textwidth]{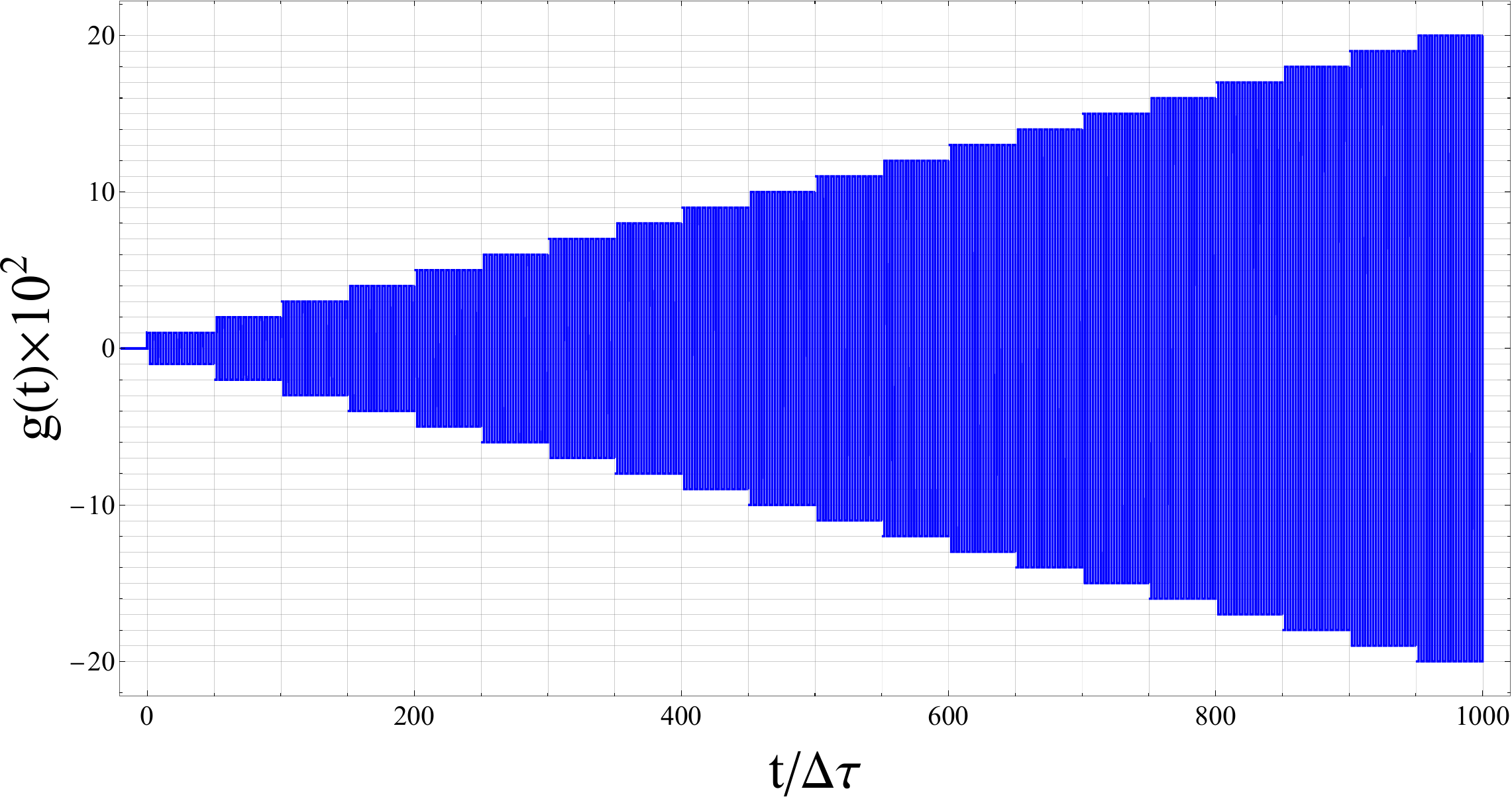}
  \caption{Modulation of the magnetization $g(t)$ for an arrangement of 20 PTCs in time with $50$ modulation periods, each PTC has an increasing amplitude of gyrotropy by steps of $0.01$. }
  \label{ValidationTimeLikeCFJED}
\end{figure}

\section{8. Validation of time-dependent effective axion field with transfer-matrix method}

To complement the effective medium picture in the previous section, we describe here the same system via transfer matrix approach. To that end, we examine the time slab composed of $N_b=20$ photonic time crystals, each of them corresponds to the periodic modulation of gyrotropy.

In turn, each photonic time crystal includes $50$ unit cells with subwavelength thickness $c\Delta \tau/\lambda \ll 1$, with each unit cell consisting of two temporal layers with the same magnitude but opposite magnetization direction. The total magnetization is constant within each photonic crystal. The first PTC starts with $g_0=0.01$, the second one with $g_0=0.02$, and so on. Such modulation of magnetization gives rise to the effective time-dependent axion field $\chi(t)= b_0 t$, with 
\begin{align}
    b_0 & \approx \frac{\pi}{2} \,\frac{0.01}{50} \, \frac{c}{\lambda}.
\label{effectiveweylfield}\end{align}

 We verify the emergence of time-dependent axion field by comparing the optical properties of a pure-time WSM slab with $b_0$ given by  Eq.~(\ref{effectiveweylfield}), and the arrangement of 20 PTCs using the transfer-matrix method. The full pattern of magnetization modulation is depicted in Fig.~\ref{ValidationTimeLikeCFJED}. Next, we plot reflection and transmission coefficients in terms of the period-to-wavelength ratio $\xi= c \Delta \tau /\lambda$. The total thickness of the time slab is $\texttt{T} = 1000 \Delta\tau$, so the plots are also proportional to the total thickness-to-wavelength ratio $c \texttt{T}/\lambda$. For simplicity, we disregard the second-order correction to the effective permeability, which now becomes time-dependent.

\begin{figure}[h!]
\centering
\includegraphics[width=0.98 \textwidth]{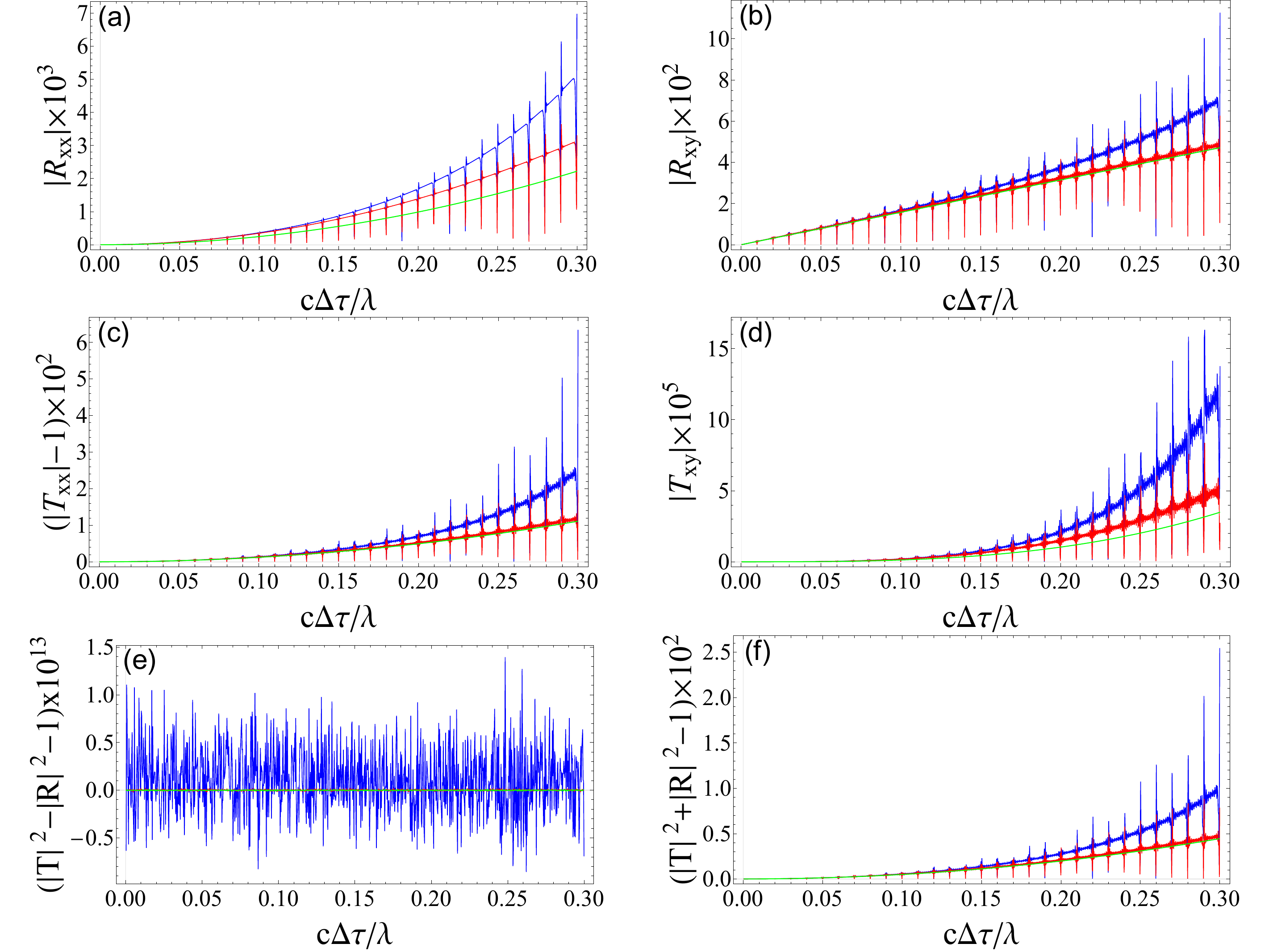}
\caption{Validation of the effective time-dependent axion response. (a-f) Blue lines correspond to the transfer-matrix method applied to the multilayered structure, red lines to the effective axion medium approximation, and the green lines represent the analytical solution for a pure-time WSM slab with $b_0$ as in Eq. (\ref{effectiveweylfield}). Panels (a,b) correspond to the co- and cross-polarized reflection, respectively, panels (c,d) show the co- and cross-polarized transmission, respectively, panel (e) displays the conservation of momentum $|T|^2-|R|^2 = 1$, and panel (f) demonstrates that energy is not conserved $|T|^2+|R|^2\neq 1$. We have considered an arrangement of 20 pure-time axion slabs with increasing gyrotropy from $0.01$ to $0.2$ in steps of $0.01$.}
\label{ValidationPureTimeLinearChi}
\end{figure}

In Fig.~\ref{ValidationPureTimeLinearChi}, we validate the description of the time slab in terms of time-dependent effective axion field.  It is evident that for rapid modulation, $\xi \ll 1$, the transmission and reflection coefficients for the real photonic time crystal arrangement (blue lines), the corresponding effective axion response (red lines), and the ideal pure-time Weyl semimetal slab (green lines) match very closely. Discrepancies between the effective axion response and the WSM arise due to second-order corrections in $\xi$ to the effective permeability, which mostly affect co-polarized reflection and co-polarized transmission. As expected, there is no energy conservation $|T|^2+|R|^2 \neq 1$, and there is a conservation of momentum $|T|^2-|R|^2 = 1$. 

It is straightforward to see that as $\xi$ takes values larger than $0.10$ the effective medium solution starts to deviate from the real photonic time crystal arrangement. Note that the agreement between the pure-time Weyl semimetal and the effective medium description will be better if we consider that the Weyl semimetal also has a gradient of permeability of the same form as suggested by the effective medium description.

\section{9. Reflection and transmission for an axion time-slab with nonzero magnetization}

In this section, we calculate the transmission and reflection coefficients of light propagating through a magnetized axion slab of thickness $\texttt{T}$. Physically this corresponds to the situation when the effective axion field and static magnetization are switched on for some finite time.

Fist we consider the temporal boundary between two gyrotropic axion slabs and use the temporal boundary conditions, Eqs. (\ref{eq:BC in Temp boundary for B1}),(\ref{eq:BC in Temp boundary for D1}), to obtain the transfer matrix across this time boundary. Since the axion field only introduces changes at the boundary, the propagation inside an axion medium with nonzero magnetization has the same form when propagating in a gyrotropic medium which we have studied in the previous section. Therefore the fields in a medium $\alpha=1,2$ can be written as

\begin{equation}
    \mathbf{B}_{\alpha,\eta} = B_{\alpha,\eta}\; e^{-i \omega_{\alpha,\eta} t} \hat{{v}}_\eta + B'_{\alpha,\eta} \; e^{i \omega_{\alpha,\eta} t} \hat{{v}}_\eta
\end{equation}
and
\begin{equation}
    \mathbf{D}_{\alpha,\eta} = \xi_{\alpha,\eta} B_{\alpha,\eta}\; e^{-i \omega_{\alpha,\eta} t} \hat{{v}}_\eta - \xi_{\alpha,\eta} B'_{\alpha,\eta}\; e^{i \omega_{\alpha,\eta} t} \hat{{v}}_\eta,
\end{equation}
    where $B_{\alpha,\eta}$ and $B'_{\alpha,\eta}$ are the amplitudes of the forward and backward waves, respectively, and 

\begin{align}
    \omega_{\alpha,\eta} & = \frac{c k}{ \sqrt{\varepsilon_\alpha \mu_{\alpha, \eta}}}, & \xi_{\alpha,\eta} & = i\eta \sqrt{\frac{\varepsilon_\alpha}{\mu_{\alpha,\eta}}}, & \hat{v}_\eta & = \frac{1}{\sqrt{2}}
    \begin{pmatrix}
        1 \\
        i \eta \\
        0 
    \end{pmatrix}, & \mu_{\alpha, \eta} & = \frac{\mu_\alpha^2}{\mu_\alpha - \eta g_\alpha}.
\end{align}
Using the boundary conditions in Eqs. (\ref{eq:BC in Temp boundary for B1}),(\ref{eq:BC in Temp boundary for D1}) we have

\begin{align}
    B_{1,\eta} + B'_{1,\eta} & = B_{2,\eta} + B'_{2,\eta} \label{boundaryaxiongyrotropictime1} \\
    \xi_{1,\eta} B_{1,\eta} -  \xi_{1,\eta} B'_{1,\eta} & =  \left(\xi_{2,\eta} - \chi\right)  B_{2,\eta} - \left(\xi_{2,\eta} + \chi\right)  B'_{2,\eta} \label{boundaryaxiongyrotropictime2}
\end{align}
with $\chi = \chi_1 -\chi_2$. The $2\times 2$ transfer matrix equation for a time-boundary of two different magnetized axion media is
\begin{align}
    \begin{pmatrix}
        B_{2,\eta}\\
        B'_{2,\eta}
    \end{pmatrix} & = M^\eta_{12} 
    \begin{pmatrix}
        B_{1,\eta}\\
        B'_{1,\eta}
    \end{pmatrix}, & M^\eta_{12} & = \frac{1}{2}
    \begin{pmatrix}
        1 + \dfrac{\xi_{1,\eta} + \chi}{\xi_{2,\eta}} &  1 - \dfrac{\xi_{1,\eta} -\chi}{\xi_{2,\eta}} \\
        1 - \dfrac{\xi_{1,\eta} + \chi}{\xi_{2,\eta}} &  1 + \dfrac{\xi_{1,\eta} - \chi}{\xi_{2,\eta}}
    \end{pmatrix}.
\end{align}
We use these temporal boundaries to write the full transfer matrix describing the incidence on a magnetized axion slab $(\chi_2=\chi, g_2 = g)$ from a conventional non-magnetized medium $(\chi_1 = g_1 = 0)$
\begin{align}
    M^\eta_{\text{AG-slab}} & = M^\eta_{2\rightarrow 1}(t=\texttt{T}) M_{\text{bulk}}(\texttt{T}) M^\eta_{1 \rightarrow 2} (t=0), &  M_{\text{bulk}}(\texttt{T}) & = \begin{pmatrix}
       e^{-i \omega_{2,\eta} \texttt{T}} & 0 \\
        0 & e^{i \omega_{2,\eta} \texttt{T}}
    \end{pmatrix},
\end{align}
and
\begin{align}
    M^\eta_{1 \rightarrow 2}(t=0) & = \frac{1}{2}
    \begin{pmatrix}
        1 + \dfrac{\xi_{1,\eta} - \chi}{\xi_{2,\eta}} &  1 - \dfrac{\xi_{1,\eta} +\chi}{\xi_{2,\eta}} \\
        1 - \dfrac{\xi_{1,\eta} - \chi}{\xi_{2,\eta}} &  1 + \dfrac{\xi_{1,\eta} + \chi}{\xi_{2,\eta}}
    \end{pmatrix}, &
    M^\eta_{2 \rightarrow 1}(t=\texttt{T}) & = \frac{1}{2}
    \begin{pmatrix}
        1 + \dfrac{\xi_{2,\eta} + \chi}{\xi_{1,\eta}} &  1 - \dfrac{\xi_{2,\eta} -\chi}{\xi_{1,\eta}} \\
        1 - \dfrac{\xi_{2,\eta} + \chi}{\xi_{1,\eta}} &  1 + \dfrac{\xi_{2,\eta} - \chi}{\xi_{1,\eta}}
    \end{pmatrix},
\end{align}
with $\xi_{1,\eta} = i \eta \sqrt{\varepsilon_0/\mu_0}$, $\xi_{2,\eta} = i \eta \sqrt{\varepsilon_0/\mu_{2,\eta}}$, and  $\mu_{2,\eta} = \mu_0^2 / (\mu_0 - \eta g)$. The components $M_{11}^\eta$ and $M_{21}^\eta$ are 

\begin{equation}
    M_{11}^\eta = \cos (\omega_\eta\texttt{T}) - i \frac{Y^2 + Y_\eta^2 + \chi^2}{2 Y Y_\eta} \sin(\omega_\eta\texttt{T}),
\end{equation}

\begin{equation}
    M_{21}^\eta = - i \frac{Y^2 - Y_\eta^2 - \chi^2 + 2 i \eta Y \chi}{2 Y Y_\eta} \sin(\omega_\eta\texttt{T}),
\end{equation}
Therefore, by using Eqs. (\ref{T:photoniccrystal}), (\ref{R:photoniccrystal}) we obtain the final results for cross-polarized transmission and reflection

\begin{align}
    T_{xy} \hphantom{-} = -T_{yx} & =  \frac{1}{2 i }\left[ \cos(\omega_+\texttt{T}) - \cos(\omega_-\texttt{T}) - \frac{i}{2} \left(\frac{Y^2 + Y_+^2 + \chi^2}{Y Y_+}\sin(\omega_+\texttt{T}) - \frac{Y^2 + Y_-^2 + \chi^2}{Y Y_-}\sin(\omega_-\texttt{T}) \right)\right] \label{Txy:puretimemagnetizedaxionslab}\\
    R_{xy} \hphantom{-} = - R_{yx} & = -\frac{1}{4 }\left[ \frac{Y^2 - Y_+^2 - \chi^2 + 2 i Y \chi}{Y Y_+}\sin(\omega_+\texttt{T}) - \frac{Y^2 - Y_-^2 - \chi^2 - 2 i Y \chi}{Y Y_-}\sin(\omega_-\texttt{T}) \right] \label{Rxy:puretimemagnetizedaxionslab}
\end{align}
with
\begin{align}
    \omega_{\eta} & = \frac{c k}{ \sqrt{\varepsilon \mu_{\eta}}}, & Y & = \sqrt{\frac{\varepsilon}{\mu}} , &Y_{\eta} & = \sqrt{\frac{\varepsilon}{\mu_{\eta}}}, & \mu_{ \eta} &= \frac{\mu^2}{\mu - \eta g}.
\end{align}
In the limit $g=0$, $\omega_+=\omega_-$ and $Y_+=Y_-=Y$. Hence, cross-polarized transmission vanishes, while cross-polarized reflection is consistent with Eq.~(\ref{Rxy:puretimeaxionslab}) for non-magnetized axion slab.

Another important limit is the situation when both axion field $\chi$ and gyrotropy $g$ are small having the same order of magnitude. Expanding the expressions above up to the first order in small parameters, we have 
\begin{align}
    \omega_{\eta} & \approx \frac{c k}{ \sqrt{\varepsilon \mu }}\Big( 1-\eta \tilde{\alpha}\Big), & Y_{\eta} & = \sqrt{\frac{\varepsilon}{\mu}} \Big( 1-\eta \tilde{\alpha}\Big), & \mu_{ \eta} & = \mu + \eta g, & \tilde{\alpha} & =\frac{1}{2} \frac{g}{\mu}
\end{align}
and by setting $\omega = ck /\sqrt{\varepsilon \mu}$, we can approximate the reflection and transmission coefficients to first order of $g$ and $\chi$ as

\begin{align}
    T_{xy} \hphantom{-}  & \approx  \tilde{\alpha}\omega   \texttt{T} \exp\skob{i \omega \texttt{T}} = \frac{1}{2} g \frac{c k}{\sqrt{\varepsilon \mu^3}} \texttt{T} \exp\skob{i\omega\texttt{T}}  \label{Txy:puretimemagnetizedaxionslablimit}\\
    R_{xy} \hphantom{-} & \approx -\Big( \frac{\mu}{\varepsilon} \tilde{\alpha} + i \chi \sqrt{\frac{\mu}{\varepsilon}} \Big) \sin(\omega \texttt{T}) = -\Big( \frac{g}{2\varepsilon}  + i \chi \sqrt{\frac{\mu}{\varepsilon}} \Big) \sin(\omega \texttt{T})\label{Rxy:puretimemagnetizedaxionslablimit}
\end{align}

\section{10. Numerical simulation for the arbitrary magnetization modulation}

Previously, we introduced transfer matrix method in time domain to analyze the properties of a medium with a stepwise modulation of gyrotropy. In this section, we outline more general yet more computationally demanding method to assess the case of arbitrary magnetization modulation. We consider gyrotropic medium with the following material parameters
\begin{align}\label{eq: gyro mu mat par}
\hat{\mu} & = \begin{pmatrix}
 \mu& i g(t) &  0\\
-ig(t) & \mu &0  \\
 0&0  &\mu  \\
\end{pmatrix}, & \hat{\eps} & =\eps.
\end{align}
Here, $g(t)$ is an arbitrary function that satisfies the only requirement $\lim_{t\to\pm\infty}g(t) =0$, i.e. modulation of magnetization finishes at some point.
%
Since the translational invariance of the system is preserved and we consider the waves propagating in $z$ direction, spatial dependence of the fields can be described via the factor $e^{ikz}$:
\begin{equation}\label{eq:sup: plane wave}
    \vc{B}(\vc{r},t) = e^{ik_zz} \,\vc{B}(t)
\end{equation}
For the above anzats, wave equation takes the following form:
\begin{equation}\label{eq:sup: wave eq with g(t)}
    \frac{\partial^2}{\partial t^2}\vc{B}+\frac{c^2k^2}{\eps\mu}\skob{1+i\mu g(t)\vc{e}_z^{\times}}\vc{B} =0.
\end{equation}
Now, we introduce a circular polarization basis, defined as follows 
\begin{equation}\label{eq:sup: circ polar}
    \vc{e}_\pm = \vc{e}_x \pm \vc{e}_y, \quad \vc{e}_z \times \vc{e}_\pm=\mp i \; \vc{e}_\pm.
\end{equation}
In this basis, \eqref{eq:sup: circ polar}, the wave equation ~\eqref{eq:sup: wave eq with g(t)} is simplified and takes the following form
\begin{equation}\label{sup:diff}
    \frac{\partial^2}{\partial t^2}B_\pm(t)+\frac{c^2k^2}{\eps\mu}\skob{1\;\pm\;\mu g(t)}B_\pm(t) = 0.
\end{equation}
Now we can solve the equations for the amplitudes $B_+$ and $B_-$ independently. We consider rapid non-uniform modulation of magnetization that starts at $t=0$ and finishes at $t=t_2$. The initial magnetic field before the start of modulation corresponds to the plane wave:
\begin{equation}
    \vc{B}(t<0) = B_0\;e^{-i\o_0t}\;\vc{e}_x, \quad \o_0 = \frac{kc}{\sqrt{\eps\mu}}
\end{equation}
This yields the following initial conditions for $B_\pm$:
\begin{equation}\label{eq:sup: initial conditions}
\begin{aligned}
    & B_+(0)=\frac{B_0}{2},& & \frac{\partial}{\partial t}B_+(0) = -i\frac{\o_0}{2}B_0  \\
    & B_-(0)=\frac{B_0}{2}, & & \frac{\partial}{\partial t}B_-(0) = -i\frac{\o_0}{2}B_0 
\end{aligned}
\end{equation}
Now we have a complete system of differential equations~\eqref{sup:diff} with initial conditions ~\eqref{eq:sup: initial conditions} which can be solved numerically.

After the modulation finishes, the magnetic field can vary in time only in a way consistent with the dispersion equation and thus
\begin{equation}
    \vc{B}(t>t_2) = \vc{B}^{(+)}\;e^{-i\o_0t} + \vc{B}^{(-)}\;e^{+i\o_0t},
\end{equation}
where $\vc{B}^{(+)}$ represents time-transmitted wave with the same phase advance as the initial field, and $\vc{B}^{(-)}$ represents time-reflected wave with the flipped phase. Transmission and reflection coefficients are generally matrices defined as:
\begin{equation}\label{eq:sup: R and T definitions}
    \vc{B}^{(+)}=\hat{T}\;\vc{B}(t=0), \quad
    \vc{B}^{(-)}=\hat{R}\;\vc{B}(t=0).
\end{equation}
Those matrices can be recovered from the solutions of Eq.~\eqref{eq:sup: wave eq with g(t)} at $t_f=t_2$. From the numerical solution, we obtain $\vc{B}(t_f)$ and $\frac{\partial}{\partial t}\vc{B}(t_f)$. Then, we can express the $\vc{B}^{(\pm)}$ fields in terms of the magnetic field, $\vc{B}$, and its time derivative at $t=t_f$ as:
\begin{equation}
    \begin{aligned}
        \vc{B}^{(+)} + \vc{B}^{(-)} & = \vc{B}(t_f),\\
        -i\o_0\vc{B}^{(+)} +i\o_0 \vc{B}^{(-)} & = \frac{\partial}{\partial t}\vc{B}(t_f),
    \end{aligned}
\end{equation}
where we assume for simplicity that $e^{-i\omega_0 t_f}=1$. From these equations, we obtain:
\begin{equation}
    \begin{aligned}
        & \vc{B}^{(+)} = \frac{1}{2}\skob{\vc{B}(t_f)+\frac{i}{\o_0}\frac{\partial}{\partial t}\vc{B}(t_f)},\\
        & \vc{B}^{(-)} = \frac{1}{2}\skob{\vc{B}(t_f)-\frac{i}{\o_0}\frac{\partial}{\partial t}\vc{B}(t_f)}.
    \end{aligned}
\end{equation}
By substituting the expressions for $\vc{B}(t_f)$ in terms of $B_\pm$ we have following expressions:
\begin{equation}
\begin{aligned}
& \vc{B}^{(+)} = \frac{1}{2}\skob{ B_+(t_f)\;\vc{e}_+ + B_-(t_f) \; \vc{e}_- }
    +\frac{i}{2\o_0}\skob{\frac{\partial}{\partial t}B_+(t_f)\;\vc{e}_+ + \frac{\partial}{\partial t}B_-(t_f) \;\vc{e}_- },\\
& \vc{B}^{(-)} = \frac{1}{2}\skob{ B_+(t_f)\;\vc{e}_+ + B_-(t_f) \; \vc{e}_- }
    -\frac{i}{2\o_0}\skob{\frac{\partial}{\partial t}B_+(t_f)\;\vc{e}_+ + \frac{\partial}{\partial t}B_-(t_f) \;\vc{e}_-}.
\end{aligned}
\end{equation}
From these expressions and Eq.~\eqref{eq:sup: R and T definitions}, we retrieve the coefficients of the transmission and reflection matrices.
\begin{align}
& T_{xx}\;=\; \frac{1}{2B_0}\skob{ B_+(t_f)+ B_-(t_f)}
    +\frac{i}{2\o_0B_0}\skob{\frac{\partial}{\partial t}B_+(t_f) + \frac{\partial}{\partial t}B_-(t_f)} \\
&  T_{yx}\;=\; \frac{i}{2B_0}\skob{ B_+(t_f) - B_-(t_f)}
    -\frac{1}{2\o_0 B_0}\skob{\frac{\partial}{\partial t}B_+(t_f) - \frac{\partial}{\partial t}B_-(t_f)} \\
& R_{xx}\;=\; \frac{1}{2B_0}\skob{ B_+(t_f)+ B_-(t_f)}
    -\frac{i}{2\o_0\;B_0}\skob{\frac{\partial}{\partial t}B_+(t_f) + \frac{\partial}{\partial t}B_-(t_f)} \\
&  R_{yx}\;=\; \frac{i}{2B_0}\skob{ B_+(t_f) - B_-(t_f)}
    +\frac{1}{2\o_0 B_0}\skob{\frac{\partial}{\partial t}B_+(t_f) - \frac{\partial}{\partial t}B_-(t_f)} 
\end{align}

We use this approach to simulate the effect of non-periodic modulation of magnetization in the manuscript main text.

\section{11. Retrieving average magnetization and effective axion field from experimental data}

Below, we demonstrate how the dynamic axion fields induced by the time-varying magnetization can be useful in mapping ultrafast magnetization dynamics. In this section, we provide the details of calculations behind Fig.~3 in the main text. In our analysis we assume that the magnetization of the structure varies according to the law
\begin{equation}\label{eq:MagnetizationDynamics}
    g(t)=g_0\,e^{-t/\tau}\,\sin\left(\Omega t+\varphi\right)\:,
\end{equation}
see Fig.~3(b) in the main text. Physically, this corresponds to the decaying oscillations of spins triggered by the powerful pump pulse. Here, $\varphi$ is the initial phase of the oscillations, $\tau$ is the relaxation time, while $\Omega$ is the frequency of the oscillations. We also compute time-reflection and time-transmission coefficients for this geometry as outlined in the previous section [Fig.~3(c,d) of the main text].

In this setting the time-averaged magnetization is nonzero and reads:
\begin{equation}\label{eq:Time-averaged}
    \left<g\right>\equiv \frac{1}{\tau}\,\int\limits_0^{\infty}\,g(t)\,dt=g_0\,\frac{\Omega\,\tau\,\cos\varphi+\sin\varphi}{1+\Omega^2\,\tau^2}\:,
\end{equation}
Specifically, when the relaxation time is much larger than the oscillation period $T=2\pi/\Omega$, i.e. magnetization oscillates multiple times before the decay, the average magnetization is roughly proportional to $\cos\varphi$.



At the same time, the oscillations of magnetization induce the effective axion field which exhibits a similar dependence on phase $\varphi$. Despite that, the two contributions can be distinguished by simultaneously measuring cross-polarized reflection and cross-polarized transmission from the structure. The analysis below shows that the cross-polarized transmission $t_{yx}$ is defined only by the average magnetization and is not affected by the effective axion field, while the cross-polarized reflection $r_{yx}$ depends on both factors simultaneously.

Now we assume that the reflection and transmission coefficients are known either from the experiment (as in the potential applications of our technique) or from numerical simulations (as in our case). As we demonstrate below, using this data, it is possible to retrieve the information on ultrafast magnetization oscillations. To that end, we approximate the time-varying magnetization by the time slab of thickness $\tau_{\text{eff}}$ with the effective axion response $\chi_{\text{eff}}$ and effective gyrotropy $g_{\text{eff}}$ which in turn depend on the phase of magnetization oscillations. We assume $\chi_{\text{eff}}$ and $g_{\text{eff}}$ to be small enough and set $\tau_{\text{eff}} \propto \tau$.

First we make use of our analytical solution for such geometry, which predicts non-zero cross-polarized reflection and transmission, Eqs.~\eqref{Txy:puretimemagnetizedaxionslablimit},\eqref{Rxy:puretimemagnetizedaxionslablimit}. 
Keeping only the first powers of small parameters $\chi_{\text{eff}}$ and $g_{\text{eff}}$, and setting $\eps=\mu=1$, we recover
\begin{align}
&  T_{xy}\;=\; \frac{1 }{2}\;g_{\text{eff}}\;\omega \;\tau_{\text{eff}}\;\exp\skob{i\;\omega\;\tau_{\text{eff}}},\label{eq: effective txy}\\
&  R_{xy}\;=\;- \skob{\frac{g_{\text{eff}}}{2}+i\;\chi_{\text{eff}}}\; \sin\skob{\omega \; \tau_{\text{eff}}}.
\label{eq: effective rxy}
\end{align}

Next we refine the value of the time slab effective thickness $\tau_{\text{eff}}$ which allows us to choose the consistent phases for cross-polarized reflection and transmission in the analytical formulas above and in the results of numerical simulations. To do that, we note that the average magnetization $g_{\text{eff}}$ should be zero for $\varphi=\pi/2$. Hence, the cross-polarized reflection in this case is purely imaginary. With this requirement and simulation parameters as in the main text, we recover $\tau_{\text{eff}}\simeq\,1.17 \tau$.


Once the effective thickness of the time slab is determined, the remaining parameters $g_{\text{eff}}$ and $\chi_{\text{eff}}$ can be readily found by fitting the data for cross-polarized time-reflection and time-transmission coefficients by the analytical formulas Eqs.~\eqref{eq: effective txy}, \eqref{eq: effective rxy}. The retrieved results are presented in Fig.~3(e,f) in the main text (orange curves) showing good agreement with the average magnetization and effective axion field calculated analytically.

Hence, we conclude that the measurement of time-reflection and time-transmission coefficients provides a tool to extract the information regarding ultrafast magnetization oscillations. We would like to highlight that the probe signal may have much lower frequency than that of the oscillating magnetization. Still, if the measurement is precise enough, i.e. cross-polarized time-reflection and time-transmission can be reliably detected, one may infer the important information about magnetization oscillations without resolving them directly.

\bibliography{Timemod}